\documentclass[a4paper,fleqn,usenatbib]{mnras}

\usepackage{newtxtext,newtxmath,amsmath}

\usepackage[T1]{fontenc}
\usepackage{ae,aecompl}

\usepackage{graphicx}	
\usepackage{amsmath}	
\usepackage{amssymb}	
\usepackage{booktabs}
\usepackage{subcaption}
\title[Star Formation, Shocks, and AGN in NGC 1068]{Separating Line Emission from Star Formation, Shocks, and AGN Ionisation in NGC 1068}

\author[J. J. D'Agostino et al.]{
Joshua J. D'Agostino$^{1,2}$\thanks{E-mail: joshua.dagostino@anu.edu.au},
Lisa J. Kewley$^{1,2}$,
Brent A. Groves$^{1,2}$, 
Anne M. Medling$^{1,3}$\thanks{Hubble Fellow},
\newauthor
Enrico Di Teodoro$^{1}$,
Michael A. Dopita$^{1,2}$\thanks{Deceased}, 
Adam D. Thomas$^{1,2}$,
Ralph S. Sutherland$^{1}$,
\newauthor
Santiago Garcia-Burillo$^{4}$ 
\\
$^{1}$Research School of Astronomy and Astrophysics, the Australian National University, Cotter Road, Weston, ACT 2611, Australia\\
$^{2}$ARC Centre of Excellence for All Sky Astrophysics in 3 Dimensions (ASTRO 3D)\\
$^{3}$Ritter Astrophysical Research Center University of Toledo Toledo, OH 43606, USA \\
$^{4}$Observatorio Astron\'omico Nacional -- OAN, Apartado 1143, 28800 Alcal\'a de Henares, Madrid, Spain\\
}

\date{Accepted XXX. Received YYY; in original form ZZZ}

\pubyear{2018}

\begin{document}
\label{firstpage}
\pagerange{\pageref{firstpage}--\pageref{lastpage}}
\maketitle

\begin{abstract}

In the optical spectra of galaxies, the separation of line emission from gas ionised by star formation and an AGN, or by star formation and shocks, are very well-understood problems. However, separating line emission between AGN and shocks has proven difficult. With the aid of a new three-dimensional diagnostic diagram, we show the simultaneous separation of line emission from star formation, shocks, and AGN in NGC 1068, and quantify the ratio of star formation, shocks, and AGN in each spaxel. The AGN, shock, and star formation luminosity distributions across the galaxy accurately align with X-ray, radio, and CO(3-2) observations, respectively. Comparisons with previous separation methods show that the shocked emission heavily mixes with the AGN emission. We also show that if the H$\alpha$ flux is to be used as a star formation rate indicator, separating line emission from as many sources as possible should be attempted to ensure accurate results.

\end{abstract}




\section{Introduction}

One of the biggest mysteries in modern astrophysics is the link between supermassive black hole (SMBH) accretion and the evolution of its host galaxy. Studies have shown SMBH mass correlates with other galaxy properties, such as the velocity dispersion \citep[$M$-$\sigma$ relation; e.g.][]{FM2000,Gebhardt2000,Tremaine2002,Gultekin2009,MM2013}, the stellar mass in the bulge \citep[$M_{BH}$-$M_{*}$ relation; e.g.][]{Magorrian1998,MH2003,Bennert2011,MM2013}, and the luminosity of the bulge \citep[$M_{BH}$-$L$; e.g.][]{MH2003,Gultekin2009,MM2013}. However, no such theory linking SMBH accretion and star formation, such as mergers, starburst-driven winds or AGN-driven outflows \citep[e.g.][]{YKS2010,Rafferty2011} have been convincing. As a result, no theoretical model has successfully been able to explain the relationship between star formation and active galactic nuclei \citep[AGN; see the review by][]{AH2012}.

Work on uncovering the link between star formation and AGN activity was explored by \citet{Kewley2001} through the use of several emission-line-ratio diagnostic diagrams given by \citet{BPT1981} and \citet{VO1987}. \citet{Kewley2001} defined a curve on each of the [O \textsc{iii}]/H$\beta$ vs [N \textsc{ii}]/H$\alpha$ \citep[commonly referred to as the `BPT diagram';][]{BPT1981}, [O \textsc{iii}]/H$\beta$ vs [S \textsc{ii}]/H$\alpha$, and [O \textsc{iii}]/H$\beta$ vs [O \textsc{i}]/H$\alpha$ diagrams to represent the theoretical maximum of line emission from star formation, derived from photoionisation modelling using the \textsc{mappings iii} code. \citet{Kauffmann2003} also defined a maximum starburst line on the BPT diagram, albeit empirically from their sample of SDSS galaxies. Contemporary work on star formation-AGN mixing treats line emission in the region below the \citet{Kauffmann2003} line on the BPT diagram as pure star formation, and line emission in the region above the \citet{Kewley2001} line is considered to be dominated by harder sources, such as AGN or shocks. The region on the BPT diagram which occupies the space between the two lines is considered to result from a mixture of star formation with an additional hard component.

The study of star formation-AGN mixing was furthered through the advent of integral field spectroscopy (IFS). Pioneered by \citet{Davies2014a,Davies2014b}, a spatially-resolved spectrum across a galaxy containing an AGN shows a very clear and smooth `mixing sequence' between star formation towards the outer of the galaxy, and AGN activity towards the centre. To calculate the relative contributions to several strong emission lines from star formation and AGN activity across the galaxy, \citet{Davies2014a,Davies2014b} empirically selected two `basis points' from the data on the BPT diagram to represent 100\% star formation and 100\% AGN activity. The relative contribution from star formation and AGN in each spaxel was then calculated by finding each spaxel's `star-forming distance', defined in \citet{Kewley2006}. \citet{TYPHOONpaper} furthered this technique by selecting five basis points through the use of photoionisation model grids for H \textsc{ii} regions and narrow-line regions (NLRs) using the \textsc{mappings v} photoionisation modelling code.

Despite improving upon the method showcased by \citet{Davies2014a,Davies2014b}, the theoretical calculation of the star formation-AGN fraction shown in \citet{TYPHOONpaper} remains incomplete. Considering the line emission from a galaxy to be the result of two sources (star formation and AGN activity) is a vast oversimplification. The BPT diagram, which has been a favoured tool for previous work on star formation-AGN mixing, fails to distinguish between star formation, AGN activity, and other sources of ionisation and excitation. In particular, shocked outflows have been shown to result from starburst galaxies \citep[e.g.][]{Heckman1987,Rupke2005}, AGN \citep[e.g.][]{Cecil2002,Rupke2011}, and galaxy mergers \citep[e.g.][]{RKD2011,RKD2014,Rupke2013}. Furthermore, shocks can result from accretion disks. Simulations and analytical calculations have shown shocks to be prevalent among the accretion disks of black holes \citep{Spruit1987,Molteni1994,SM1994} and have been observed in the accretion disk of the dwarf nova U Geminorum \citep{NB1998}. Additionally, ionisation and excitation between an AGN and shocks can be further confused through differences in observed phenomena, such as in the AGN spectral energy distribution (SED) or continuum \citep[e.g.][]{FM1982,Kraemer2009}, or even gas turbulence (particularly when using [N \textsc{ii}]$\lambda 6584$ as a diagnostic; \citealt{GS2017}). Yet on the BPT diagram, shocked line emission can be found in the same region as AGN-affected spaxels, or within the mixing sequence of the galaxy \citep[for the position of shock features on diagnostic diagrams, see][]{Rich2010,RKD2011,Kewley2013a}. Hence, when studying a mixing sequence from a galaxy, simplifying the emission into a ratio between only star formation and AGN activity can lead to misleading if not erroneous results. 

A three-component emission decomposition between star formation, shocks, and AGN activity was performed by \citet{Davies2017}, by using the BPT diagram as well as the [O \textsc{iii}]/H$\beta$ vs [S \textsc{ii}]/H$\alpha$ diagnostic diagram. Similarly to \citet{Davies2014a,Davies2014b}, the basis points to represent 100\% star formation, shock and AGN line emission were selected empirically. Despite utilising another diagnostic diagram, the [O \textsc{iii}]/H$\beta$ vs [S \textsc{ii}]/H$\alpha$ diagram is just as problematic as the BPT diagram when attempting to separate shocked line emission from star formation- and AGN-ionised line emission. 

Throughout this work, we use a new three-dimensional diagnostic diagram first described in \citet{3dletter}, which shows the separation between star formation, shocks, and AGN line emission more clearly than the BPT diagram. Further, we follow \citet{TYPHOONpaper} by using theoretical photoionisation modelling to define regions on the 3D diagnostic diagram of 100\% star formation, shocks, and AGN. In Sections~\ref{sec:s7} and~\ref{sec:1068}, we provide detail on the IFU survey data used in this paper, as well as providing information on our test case NGC 1068. The main work of this paper is found in Section~\ref{sec:work}, which includes demonstrating the 3D diagnostic diagram, as well as calculating the star formation, shock, and AGN fractions in NGC 1068. We compare our results to data in wavelengths outside of the optical range in Section~\ref{sec:wavelengths}, before summarising our findings and describing future applications of this method in Section~\ref{sec:conclusions}. 

\section{Siding Spring Southern Seyfert Spectroscopic Snapshot Survey (S7)}
\label{sec:s7}

The Siding Spring Southern Seyfert Spectroscopic Snapshot Survey (S7) is an optical IFU survey performed between 2013 and 2016. The Wide Field Spectrograph \citep[WiFeS;][]{wifes1,wifes2} was used for the S7 survey, located on the ANU 2.3m telescope at Siding Spring Observatory. WiFeS has a field-of-view of $38 \times 25$ arcsec$^{2}$ composed of a grid of $1 \times 1$ arcsec$^{2}$ spaxels. The S7 is an ideal choice for this work, due to the high spectral resolution and spatial resolution of the survey. In particular, the high spectral resolution ($R \sim 7000$ in the red, 3000 in the blue) allows the study of independent velocity components in the emission lines, providing more information on shocked regions \citep{Ho2014}. Full details of the S7, including wavelength ranges and selection criteria, can be found in \citet{DopitaS7} and~\citet{S7}.

\section{NGC 1068}
\label{sec:1068}

NGC 1068 is the prototypical Seyfert 2 \citep{OM1993} (R)SA(rs)b \citep{dV1991} galaxy,  roughly located at a distance of 12.5 Mpc \citep[e.g.][]{1068dist} in the constellation Cetus. It has an optically-thick torus, which is known to obscure the broad-line region of the galaxy \citep[e.g.][]{MA1983,AM1985,Miller1991,GB2016}. However, contemporary work has begun to resolve the circumnuclear disk and nuclear region of NGC 1068 \citep{Marinucci2016,GB2016}.

A large-scale biconical outflow with position angle ${\sim} 32^{\mathrm{o}}$ is seen from the centre of the galaxy, observed to be moving outward at velocities of ${\sim} 3000$ kms$^{-1}$ \citep{Cecil2002}. Such an outflow is believed to be the result of radiative acceleration from the central AGN \citep{Pogge1988,Cecil2002,Dopita2002b,TYPHOONpaper}. This conclusion is supported by results at radio wavelengths. Emission from dense molecular gas tracers, in particular CO(3-2), CO(6-5), HCN(4-3), HCO+ (4-3), and CS(7-6), was mapped by \citet{GB2014} using ALMA. Using these maps, \citet{GB2014} show a large outflow in all molecular gas tracers in the inner 50 - 400 pc, with a mass of $M_\mathrm{mol} = 2.7^{+0.9}_{-1.2} \times 10^7 \;M_\odot$. A tight correlation between the motions in the circumnuclear disk (CND), the radio jet, and the ionised gas outflow is also shown by \citet{GB2014}. Such a correlation suggests an AGN-driven outflow. This notion is further supported when considering the outflow rate in the circumnuclear disk. \citet{GB2014} state that the CND outflow rate of $63^{+21}_{-37} \;M_\odot$ yr$^{-1}$ is much larger than the star formation rate (SFR) calculated at the same radii in NGC 1068. Furthermore, \citet{Tacconi1994} conclude that gas at the leading edge of the bar in NGC 1068 is shocked, through observations in the near-infrared.

Young stellar clusters towards the centre of NGC 1068 have previously been identified by \citet{SB2012}, supporting the idea of nuclear star formation. However, the nuclear star formation in NGC 1068 is small, with nuclear SFR estimates of $SFR_\mathrm{nuclear} \sim 0.4 - 1.0 \;M_\odot$ yr$^{-1}$ being calculated between radii of 12 - 140 pc \citep{Davies2007,Esquej2014}. This is in stark contrast to the SFR calculated away from the nucleus, up to radii of several kpc. Surrounding the CND, \citet{Thronson1989} calculate a SFR of ${\sim} 100 \;M_\odot$ yr$^{-1}$, likely a result of the high molecular mass found in the region (${\sim} 2-6 \times 10^9 \;M_\odot$). Such a large mass is believed to have been confined to the region surrounding the CND by the bar at the centre of NGC 1068 \citep{Thronson1989}. Hence, the majority of star formation in NGC 1068 is seen in what is aptly named the `star-forming ring' towards the central region of the galaxy. A corollary to such a statement is that extensive star formation towards the outskirts of NGC 1068 is very rarely seen.

\section{Photoionisation models}
\label{sec:models}

We use photoionisation models produced with \textsc{mappings v.1} \citep{mappingsv} to aid in the separation of star formation, shock and AGN line emission. Models of H \textsc{ii} regions and NLRs are used to represent line emission from star formation, and shocks/AGN respectively. We use identical H \textsc{ii} region and NLR models to those described in \citet{TYPHOONpaper}, with the exception of using a NLR model pressure of $P/k = 2 \times 10^7 \;\mathrm{cm}^{-3}\;\mathrm{K}$, corresponding to assumptions of an initial NLR temperature of 20,000 K and initial electron density of $n = 1000\;\mathrm{cm}^{-3}$. An initial electron density of $n_e = 1000\;\mathrm{cm}^{-3}$ for the NLR of NGC 1068 is justified through the electron density distribution shown in \citet{TYPHOONpaper}, calculated using the [S \textsc{ii}] doublet ratio. An initial temperature of 20,000 K is considered to be within the typical range for NLRs \citep[e.g.][]{Woltjer1959,CC1983,Macalpine1986}. We note however that an electron temperature of ${\sim} 20,000$ K is calculated using the [O \textsc{iii}] ratio \citep[e.g.][]{Vaona2012}, and the use of different emission line ratios (e.g. the [O \textsc{ii}] or [S \textsc{ii}] ratios) may lead to a broader range of possible calculated temperatures for the NLR \citep{Taylor2003,Vaona2012}. For a detailed explanation of the parameters associated with the H \textsc{ii} region and NLR models, see \citet{TYPHOONpaper}.

%
%
%
%
%

\section{Separating line emission from star formation, shocks, and AGN}
\label{sec:work}

\subsection{The need for a new diagnostic}

Previous studies which considered separating line emission from star formation and an additional high energy component, such as shocks \citep[e.g.][]{Rich2010,RKD2011,RKD2014}, or AGN \citep[e.g.][]{Kewley2001,Kewley2006,Kewley2013a,Kewley2013b,Davies2014a,Davies2014b,TYPHOONpaper} utilised the BPT diagram most favourably, as well as other diagnostic diagrams such as the [O \textsc{iii}]/H$\beta$ vs [S \textsc{ii}]/H$\alpha$ and the [O \textsc{iii}]/H$\beta$ vs [O \textsc{i}]/H$\alpha$ diagnostic diagrams introduced by \citet{VO1987}. The radiation field produced by sources such as shocks or AGN is harder than that from star formation, leading to increased excitation of collisionally-excited emission lines, such as [N \textsc{ii}]$\lambda$6584, and [O \textsc{iii}]$\lambda$5007. Hence, separation of star formation and line emission from harder sources is straightforward on an emission line diagnostic diagram.

These diagnostic diagrams fail however, when attempting to separate line emission between shocks and AGN. Shocks and AGN can produce very similar values of the line ratios on the aformentioned diagnostic diagrams, leading to the two sources being indistinguishable if both shocks and AGN are present in a single galaxy. Furthermore, shocked spaxels on the BPT diagram in particular may coincide with spaxels along the galaxy's mixing sequence \citep[see][]{Davies2014a,Davies2014b,TYPHOONpaper}, leading to further confusion. Hence, in order to simulataneously separate line emission from star formation, shocks, and AGN activity in a single galaxy, a new diagnostic diagram must be explored.

\subsection{The 3D diagnostic diagram}

To simultaneously separate line emission from star formation, shocks, and AGN, we utilise the 3D diagnostic diagram first shown by \citet{3dletter}. We consider the 3D diagnostic diagram to be an extension of the BPT diagram, combining the BPT line ratios with distance and velocity dispersion. The information from the BPT diagram is retained by using a function of the two line ratios, defined as the emission-line-ratio (ELR) function. The ELR functional form is given in Equation~\ref{eq:elr}. The ELR function is defined in such a way that any two hardness-sensitive line ratios may be used instead of those from the BPT diagram. We have chosen the BPT line ratios in particular, as the BPT diagram has been historically the most favoured diagnostic diagram for studying the separation of star formation, and AGN or shocks. Other diagnostic line ratios such as [O \textsc{i}]/H$\alpha$ or [S \textsc{ii}]/H$\alpha$ may be easily substituted into the ELR function. The purpose of the ELR function is to simply order the spaxels on the 3D diagram in terms of their combined [O \textsc{iii}]/H$\beta$ and [N \textsc{ii}]/H$\alpha$ ratios. We emphasise that the ELR function is data-dependent, and the range of values from 0 to 1 is arbitrary. The endpoints of 0 and 1 do not represent any physical phenomena, such as 100\% star formation emission and 100\% AGN emission respectively. 

\begin{figure*}
\centering
\includegraphics[width=\textwidth]{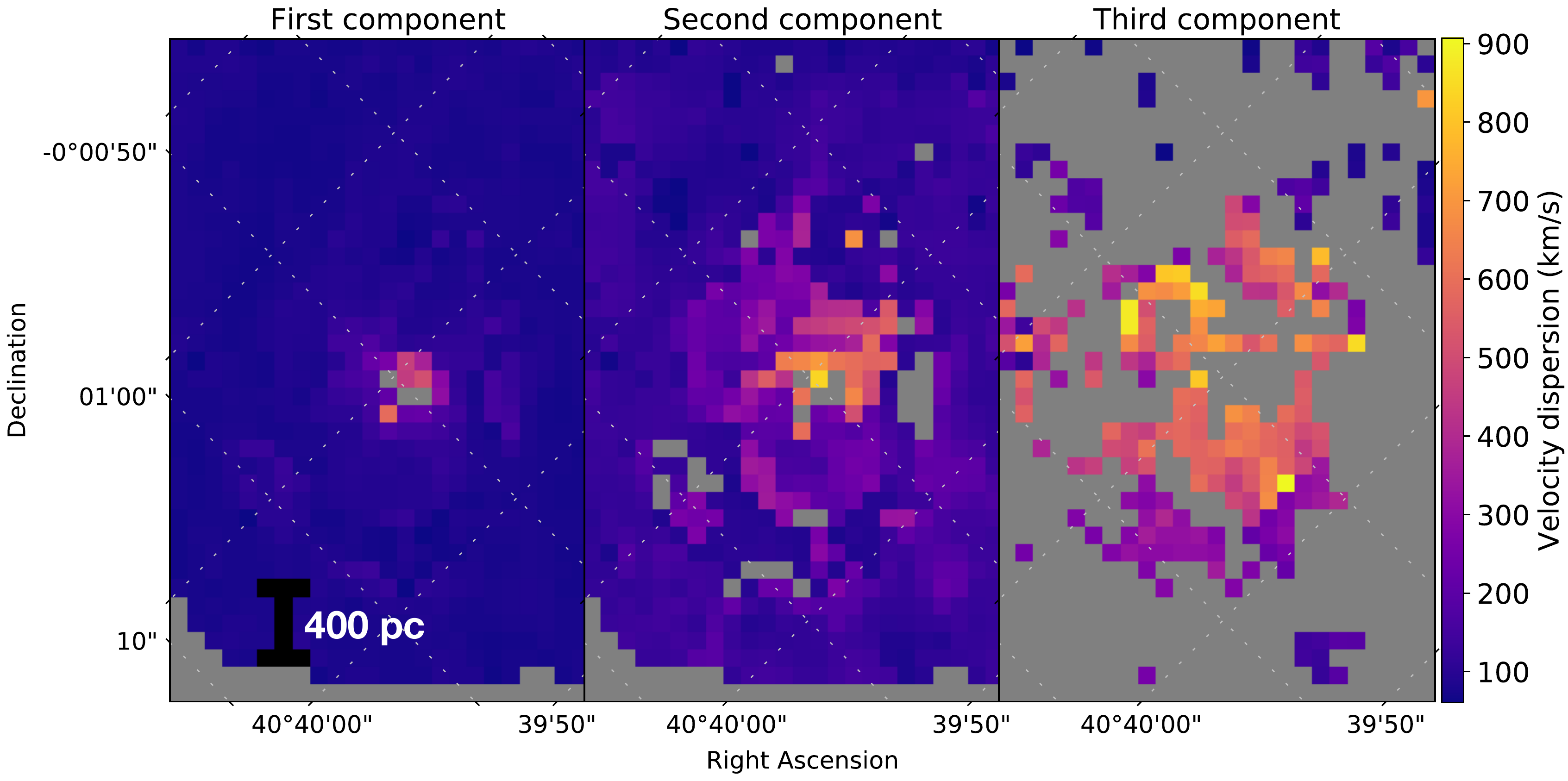}
\caption{Maps of NGC 1068 showing the distribution and amplitude of each individual velocity component in the data. Dashed lines represent grid lines of constant right ascension and declination.}
\label{fig:vdispmaps}
\end{figure*}

Prior to the data being displayed on the 3D diagnostic diagram, we first reduce and flux-calibrate the emission-line fluxes in the raw cubes using \textsc{lzifu} \citep{lzifu}. \textsc{lzifu} is capable of fitting multiple Gaussian components to each emission line, thereby resolving multiple velocity components. Once each emission line is fit with a maximum of three Gaussian components, the neural network \textsc{LZComp} \citep{Hampton2017} is used to determine the recommended number of Gaussian components for each emission line in each spaxel. Maps showing the distribution across the galaxy of each resolved velocity component in each spaxel are shown in Figure~\ref{fig:vdispmaps}. Full details of the S7 data reduction are provided in \citet{S7}. 

\begin{figure*}
\centering
\includegraphics[width=\textwidth]{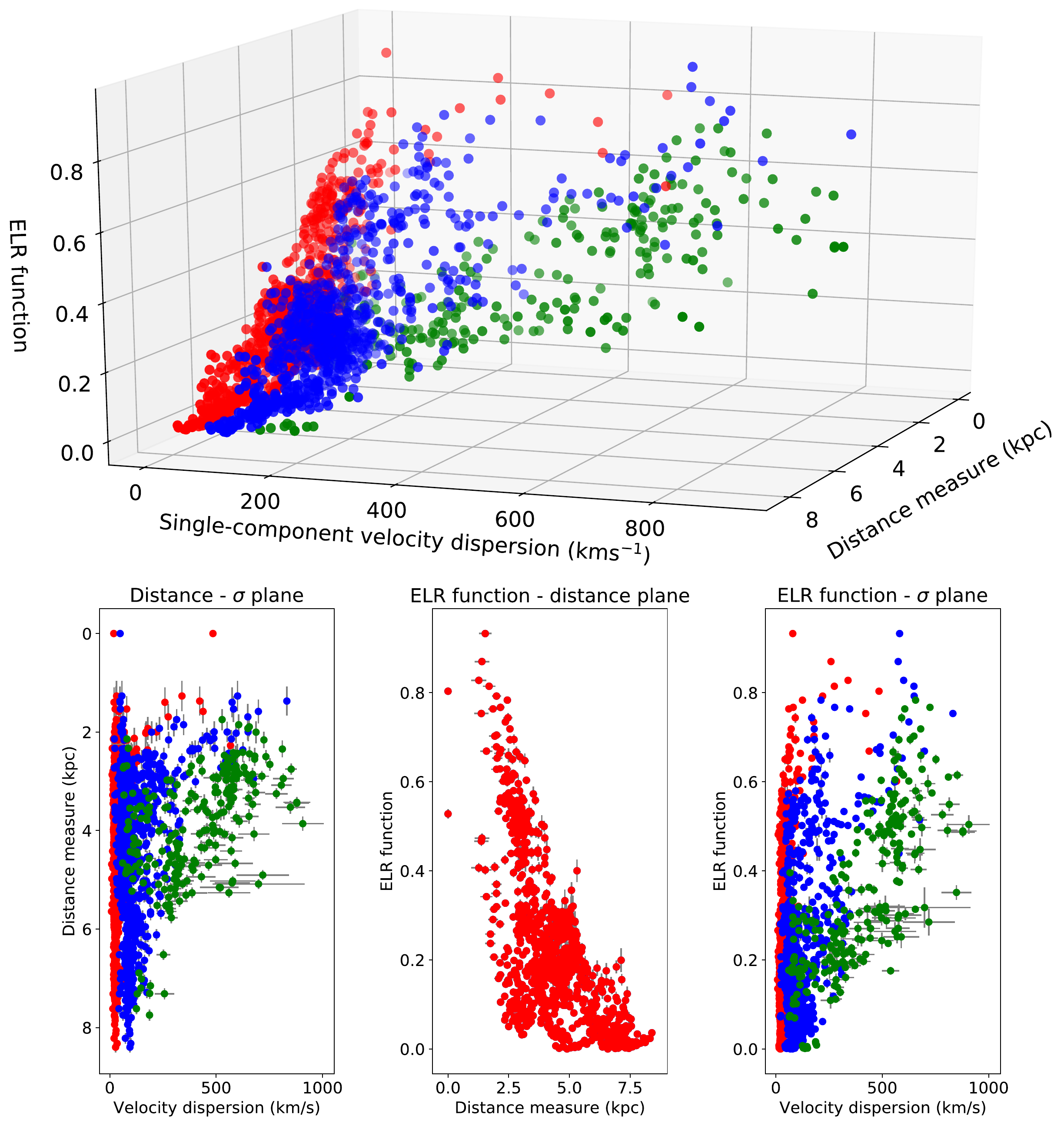}
\caption{3D diagnostic diagram showing data from NGC 1068. Red spaxels are spaxels using the first-component fits to the velocity dispersion, blue spaxels are the second-component fits, and green spaxels are the third-component fits. First-, second-, and third-component velocity components in each spaxel, if present, all use the total flux (`zeroth' component) and the distance measurement for the individual spaxel. The ELR function is given in Equation~\ref{eq:elr}. Bottom panels show the 2D projections in the distance-$\sigma$, ELR function-distance, and ELR function-$\sigma$ planes. Grey lines in each of the three panels represent the errors associated with each dimension. Errors are omitted from the 3D diagram for clarity.}
\label{fig:3ddiags}
\end{figure*}

An example of the 3D diagnostic diagram showing data from NGC 1068 is shown in Figure~\ref{fig:3ddiags}. The data points shown on the 3D diagram represent each individual velocity component in the data. For every component present in each spaxel, the velocity dispersion of the individual component (hereafter referred to as the `single-component velocity dispersion') is combined with the total flux (`zeroth' component) and the radial value for the spaxel to define the position on the figure. Full details of the 3D diagram can be found in \citet{3dletter}. Data points which contain the first-component, second-component, and third-component fits to the velocity dispersion are shown on the 3D diagram in Figure~\ref{fig:3ddiags} in red, blue, and green respectively. The first, second, and third components are numbered in order of the narrowest to broadest velocity dispersion. Hence, higher-order components will have greater values of the single-component velocity dispersion. We include the higher-order component fits to the velocity dispersion on the 3D diagram, as shocks are typically diagnosed and categorised by high velocity dispersions \citep{Rich2010,RKD2011,RKD2014,Ho2014}. All relevant emission lines are corrected for extinction, in accordance with \citet{CCM1989}. The extinction calculation from \citet{CCM1989} uses the colour excess $E(B - V)$, which itself uses the Balmer decrement H$\alpha$/H$\beta$.

\begin{equation}
\begin{aligned}
\mathrm{ELR \;function} = \frac{\mathrm{log([N \textsc{ii}]/H}\alpha) - \mathrm{min}_{\mathrm{log([N \textsc{ii}]/H}\alpha)}}{\mathrm{max}_{\mathrm{log([N \textsc{ii}]/H}\alpha)} - \mathrm{min}_{\mathrm{log([N \textsc{ii}]/H}\alpha)}}\\
 \times \frac{\mathrm{log([O \textsc{iii}]/H}\beta) - \mathrm{min}_{\mathrm{log([O \textsc{iii}]/H}\beta)}}{\mathrm{max}_{\mathrm{log([O \textsc{iii}]/H}\beta)} - \mathrm{min}_{\mathrm{log([O \textsc{iii}]/H}\beta)}}
\end{aligned}
\label{eq:elr}
\end{equation}

\subsubsection{Effects of beam smearing on the velocity dispersion}
\label{sec:beamsmearing}

The finite beam size of a telescope inevitably leads to a phenomenon known as beam smearing \citep[e.g.][]{Bosma1978,Begeman1987}. The smearing of line emission across adjacent spaxels produces broader line profiles, which ultimately can be erroneously interpreted as a higher velocity dispersion in a given spaxel. This could prove problematic, because the 3D diagram from \citet{3dletter} uses the velocity dispersion in a spaxel as a major shock diagnostic. Large velocity dispersions towards the centre of an AGN are predominantly the result of outflows from the NLR \citep[e.g.][]{Freitas2018}. Thus, the impact of beam smearing may be small, yet potentially significant nonetheless. 

\begin{figure}
\centering
\includegraphics[width=\columnwidth]{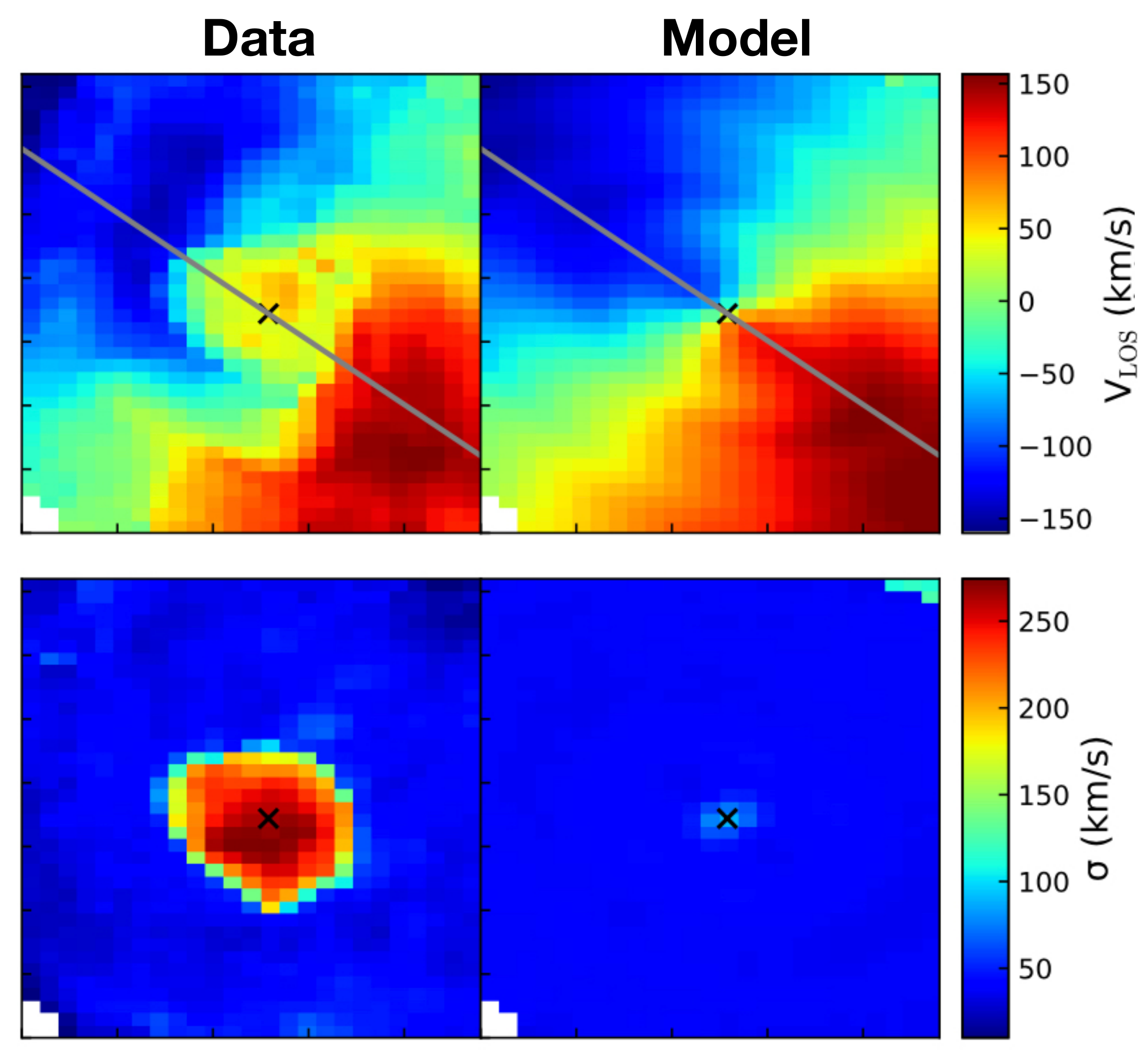}
\caption{Maps of the line-of-sight velocity ($v_\mathrm{LOS}$) and the velocity dispersion ($\sigma$) for NGC 1068 in the S7 field-of-view ($38 \times 25$ arcsec$^2$). The left-hand panels show the maps obtained from the data directly. The right-hand panels show the maps of the model produced using \textsc{$^{3\mathrm{D}}$barolo}. The centre of the galaxy is marked with an `X' in all panels.}
\label{fig:bbaroloresults}
\end{figure}

To verify that the large velocity dispersions seen in NGC 1068 are not solely the result of beam smearing, we use the code \textsc{$^{3\mathrm{D}}$barolo} \citep{Bbarolo} to recover the rotational kinematics of the galaxy. The results of using \textsc{$^{3\mathrm{D}}$barolo} are shown in Figure~\ref{fig:bbaroloresults}. The velocity field is fit using the H$\alpha$ line in the \textsc{$^{3\mathrm{D}}$barolo} model. The model panels in Figure~\ref{fig:bbaroloresults} (right-hand-side panels) show maps of the line-of-sight velocity field and velocity dispersion of the derived rotation curve of the galaxy. We are only concerned with the central region of the galaxy, because beam smearing is associated with strong velocity gradients. Strong velocity gradients are more likely found towards the centre of a galaxy due to disk rotation, where the velocity field will transition between positive and negative velocities rapidly. Figure~\ref{fig:bbaroloresults} shows that in the ideal model case, beam smearing will only increase the velocity dispersion at the centre of NGC 1068 slightly (seen as the small region of ${\sim} 100$ kms$^{-1}$ spaxels in the model velocity dispersion panel). The data however, shows a much larger region with very high velocity dispersions. This provides confidence that the high velocity dispersions at the centre of NGC 1068 are physical, and not the result of beam smearing. Furthermore, high spatial resolution images from \emph{Hubble Space Telescope (HST)} \citep[e.g.][]{Cecil2002} show high-velocity regions (up to ${\sim} 3000$ kms$^{-1}$) within the central arcsecond of NGC 1068, supporting our claim of the physical nature of these high velocity dispersions.

\subsubsection{Calculating the distance measurement in each spaxel}

\begin{figure*}
\centering
\begin{subfigure}{0.5\textwidth}
\includegraphics[width=\linewidth]{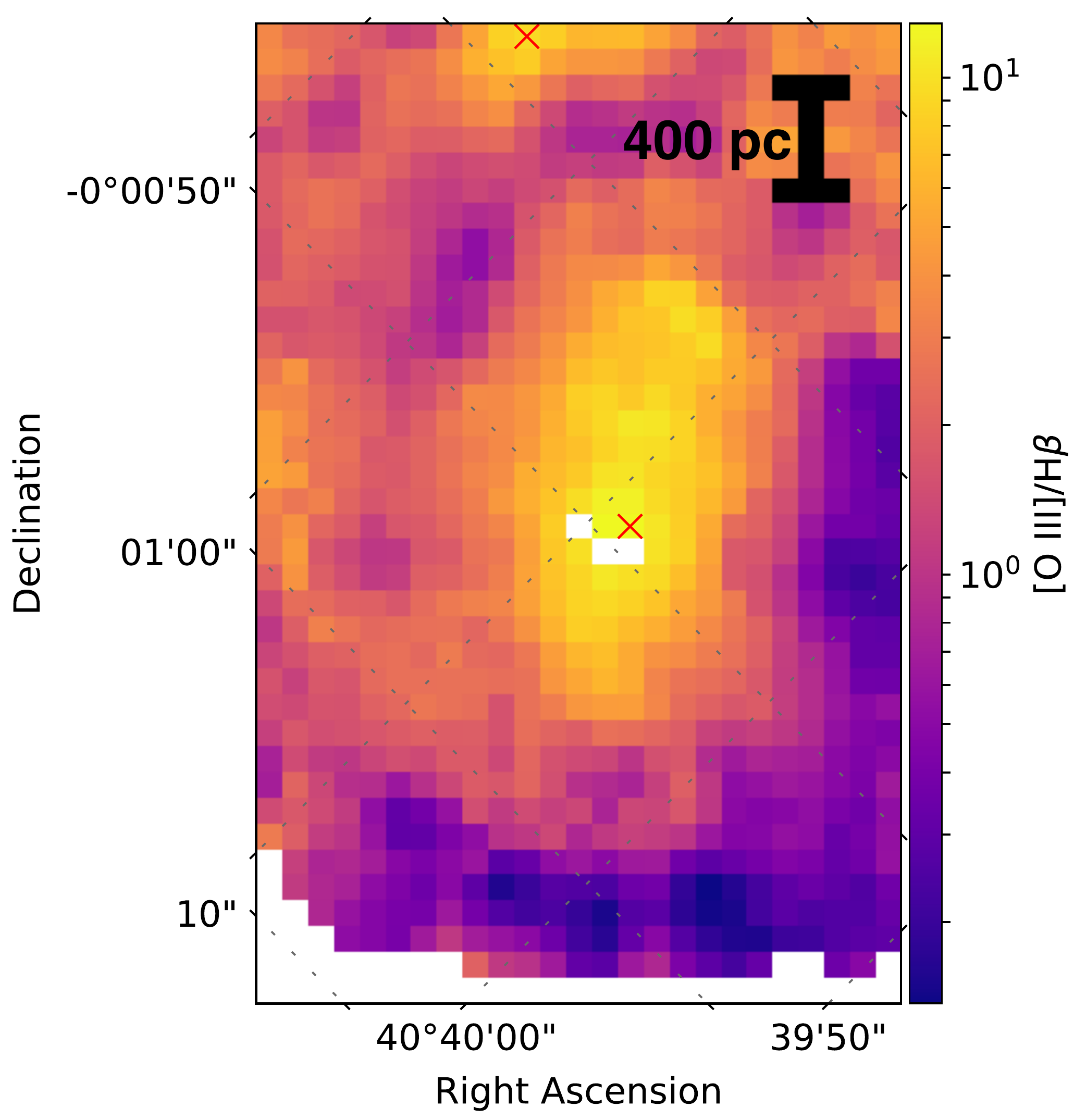}
\caption{[O \textsc{iii}]/H$\beta$ map}
\label{fig:oIIImapsa}
\end{subfigure}
\begin{subfigure}{\columnwidth}
\includegraphics[width=\linewidth]{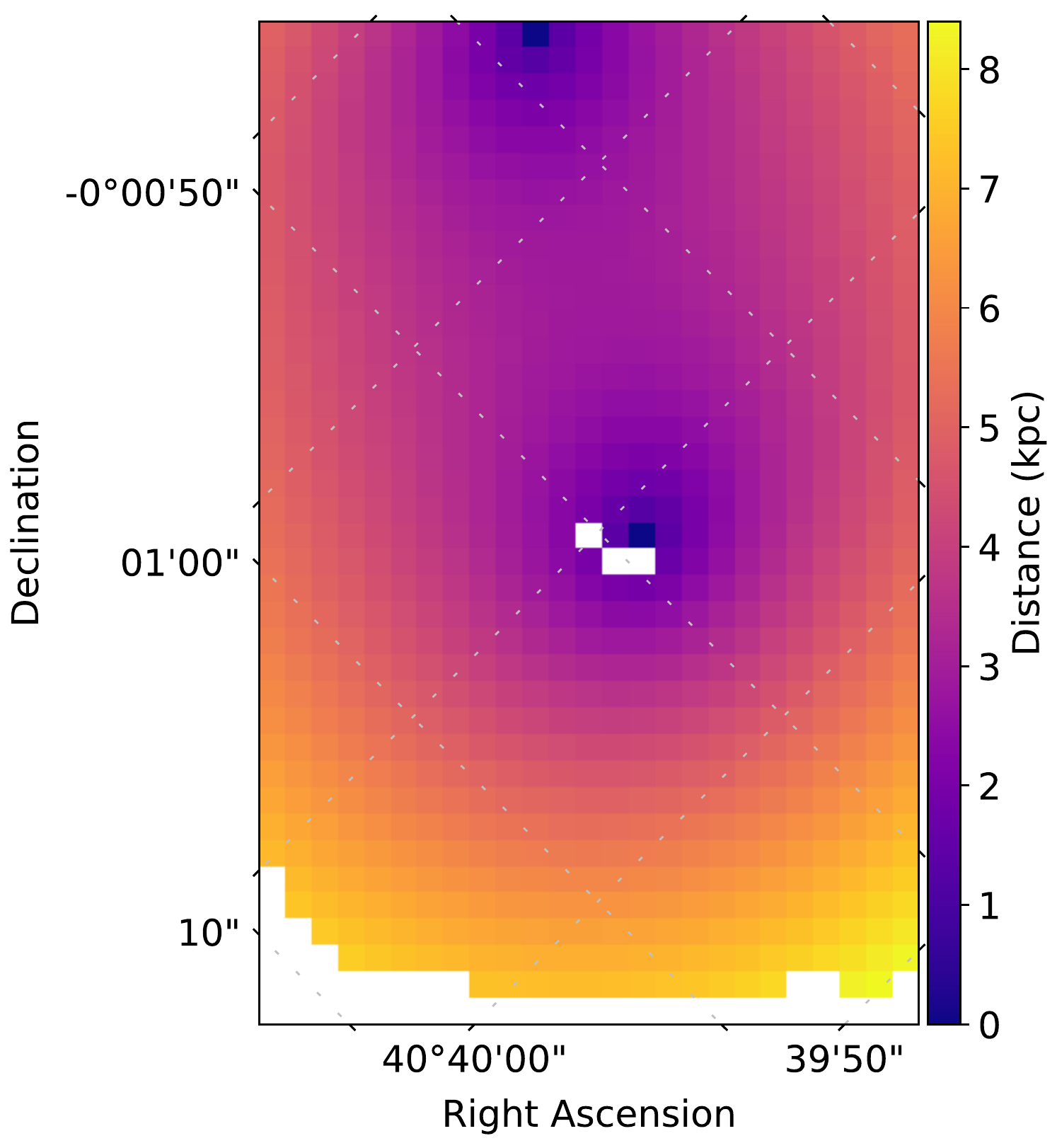}
\caption{Distance map}
\label{fig:oIIImapsb}
\end{subfigure}
\caption{[O \textsc{iii}]/H$\beta$ map in (a) and distance map in (b) for NGC 1068. The distance in each spaxel is calculated by first identifying peaks in the [O \textsc{iii}]/H$\beta$ distribution. Peaks are identified in (a) by red crosses. Dashed lines represent grid lines of constant right ascension and declination.}
\label{fig:oIIIhbmaps}
\end{figure*}

In order to assign each spaxel a distance value, we first check for local maxima in the [O \textsc{iii}]/H$\beta$ distribution across the galaxy. Local maxima in the [O \textsc{iii}]/H$\beta$ distribution are identified, and shown as red crosses in Figure~\ref{fig:oIIIhbmaps}a. Each maximum is treated as a centre, and a deprojected radius is calculated from each maximum, assuming the same values for the position angle and axis ratio for the galaxy. In the event where multiple maxima are detected (as in the case of NGC 1068), the deprojected radii calculated from each maximum are multiplied together, and then raised to the power of $1/n_\mathrm{maxima}$, where $n_\mathrm{maxima}$ is the number of maxima detected (the `geometric mean'). If only a single peak is detected in the [O \textsc{iii}]/H$\beta$ map, the peak is treated as the centre of the galaxy, and the distance measurement defaults to the deprojected galactocentric radius. A map showing the distribution of distance values across NGC 1068 is shown in Figure~\ref{fig:oIIIhbmaps}b. Hereafter, the distance measurement may simply be referred to as `distance'.

\subsection{Defining the star formation, shock, and AGN extrema}
\label{sec:extrema}

The 100\% regions in the 3D diagram indicate where emission is expected to be solely from one ionising source. Each of the 100\% star formation, shock, and AGN regions of the 3D diagram are defined by a value of the ELR function, similarly to the spaxels on the BPT diagram. We calculate the 100\% star formation, shock, and AGN ELR function values through photoionisation modelling using \textsc{mappings v.1}. 

\begin{figure}
\centering
\includegraphics[width=\columnwidth]{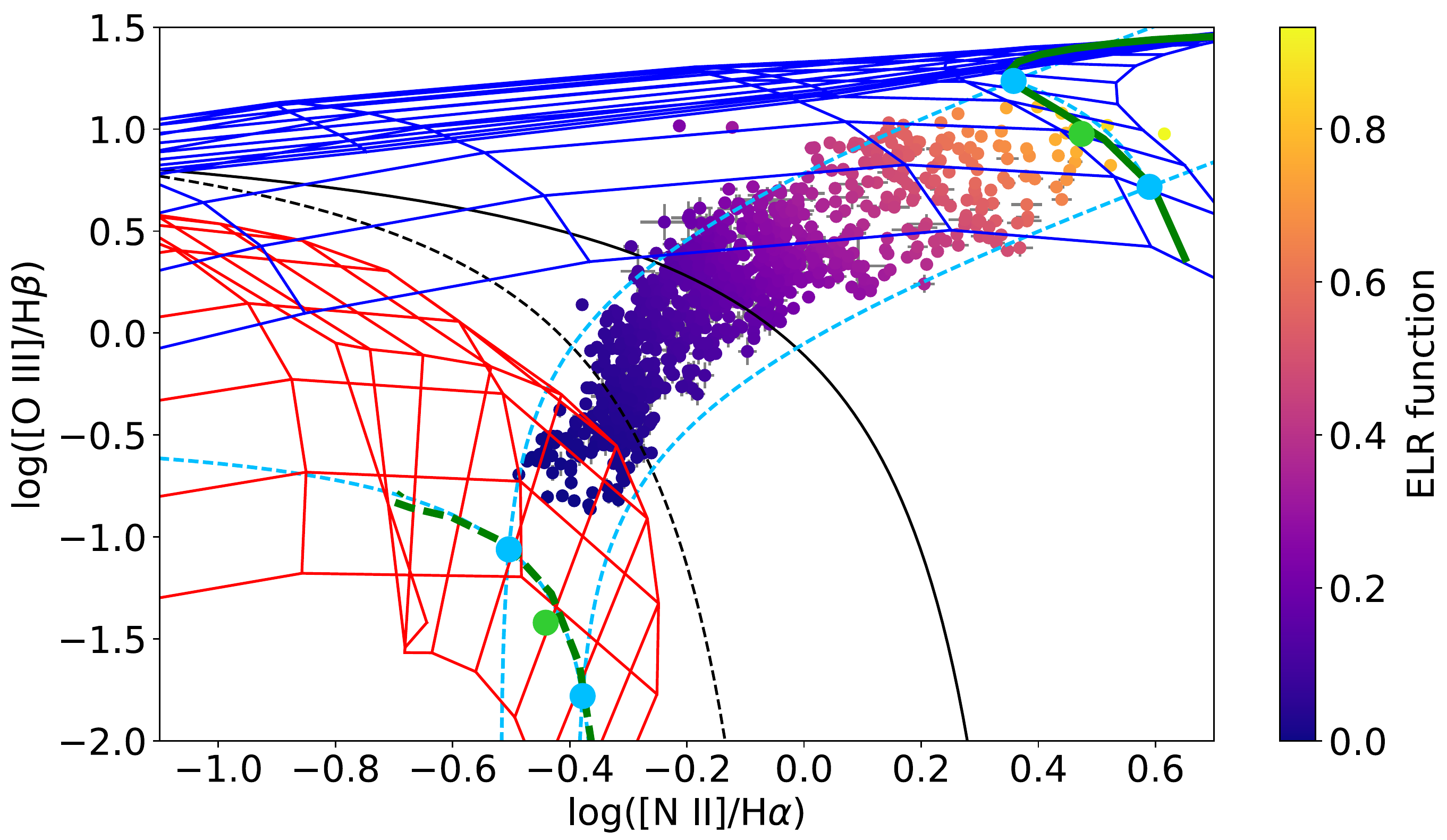}
\caption{BPT diagram of NGC 1068 with spaxels coloured to the ELR function in Equation~\ref{eq:elr}. The red grid is the H \textsc{ii} region model grid, and the blue grid is the NLR model grid, both described in Section~\ref{sec:models}. The green dashed line on the H \textsc{ii} region grid is a constructed line of constant metallicity for the outer regions of the galaxy, within the S7 field-of-view. The solid green line on the NLR model is the constructed line of constant metallicity for the centre of the galaxy. Both metallicity values are found in Table~\ref{tab:values}. The light blue points are the basis points used to calculate the star formation and AGN/shock extrema for the 3D diagnostic diagram. Light green points represent the mean value of each pair of light blue basis points, and define the 100\% star formation and AGN/shock ELR function values, also found in Table~\ref{tab:values}. The dashed black line is the \citet{Kauffmann2003} demarcation line, and the solid black line is the \citet{Kewley2001} demarcation line.}
\label{fig:bpt}
\end{figure}

The 100\% star formation and AGN emission line ratios for NGC 1068 are obtained by using the BPT diagram directly. We first construct new grid lines of constant metallicity on the H \textsc{ii} region and NLR grids, after calculating the metallicity gradient of the galaxy using the \citet{KD2002} (hereafter KD02) metallicity diagnostic. The [N \textsc{ii}]/[O \textsc{ii}] diagnostic described by KD02 is a good choice when attempting to calculate the metallicity gradient of galaxies which contain an AGN. \citet{KE2008} show that the relative contribution from an AGN has a very minor effect ($\leq 0.04$ dex at ${\sim} 15$\% AGN contribution) on the [N \textsc{ii}]/[O \textsc{ii}] ratio, in spaxels below the \citet{Kewley2001} theoretical maximum starburst line. The central and outer metallicities for NGC 1068 are shown in Table~\ref{tab:values}. The newly-constructed line of constant metallicity in the H \textsc{ii} region grid uses the outer metallicity of the galaxy, whilst the line constructed on the NLR grid uses the central metallicity. This is because the mixing sequence of the galaxy traces increasing AGN contribution from H \textsc{ii} regions outside of the AGN extended narrow-line region (ENLR) towards the AGN at the centre of the galaxy \citep[see][]{Kewley2001,Kewley2013a,Davies2014a,TYPHOONpaper}.

We select two basis points each on both the new H \textsc{ii} region and NLR grid lines, to account for the spread in ionisation parameter in the mixing sequence on the BPT diagram. The basis points are intended to signify the regions of 100\% star formation and 100\% AGN on the H \textsc{ii} region and NLR grids respectively. These basis points are located along the new lines of constant metallicity, and are shown in light blue on Figure~\ref{fig:bpt}. The mean value of the pair of basis points then determines the 100\% ELR function value for the star-forming and AGN regions. The point representing the mean value for each pair is shown in light green on Figure~\ref{fig:bpt}. We use the 100\% AGN ELR function value for the 100\% shock ELR function value also. This is because shocks may be prevalent in the NLR of galaxies, produced from sources such as the radiation pressure from accretion disk emission, and the pressure from the radio jets \citep[e.g.][and references therein]{Dopita1995,WR1999,Dopita2015}. The values of the ELR function used to represent 100\% star formation, AGN, and shock emission are given in Table~\ref{tab:values}.


\begin{table}
\centering
\begin{tabular}{| c | c |}
\toprule
Outer metallicity ($Z_\odot$) & 1.48 \\
Central metallicity ($Z_\odot$) & 2.20 \\
100\% star formation ELR function value & -0.01 \\
100\% shock/AGN ELR function value & 0.81 \\
\bottomrule
\end{tabular}
\caption{Metallicity gradient and 100\% emission line ratio values for NGC 1068. Solar metallicity ($Z_\odot$) is set to 12 + log(O/H) = 8.93 \citep{AG1989}.}
\label{tab:values}
\end{table}

\subsection{Calculating the star formation-shock-AGN fraction}

The positions of the spaxels on the 3D diagram are important in quantifying the contribution to emission from star formation, shocks, and AGN in each spaxel. Seen in Figure~\ref{fig:3ddiags}, the spread of spaxels on the 3D diagram shows two distinct and clear sequences. The first sequence contains spaxels at low velocity dispersions, and shows very little change in velocity dispersion with increasing values of the ELR function. The second sequence however displays a clear increase in velocity dispersion as the value of the ELR function increases in the spaxels. \citet{3dletter} show that the first and second sequences represent the star formation-AGN and star formation-shock mixing respectively (see \citealt{3dletter} for complete details on the diagram and its interpretation). 

Each data point on the 3D diagnostic diagram is assigned a star formation-shock-AGN fraction. The star formation-shock-AGN fraction for each data point is calculated by considering its distance along the two sequences. Points containing high ELR function values, found towards the top of the first and second sequences, are assigned high AGN and shock fractions respectively. Conversely, points with low ELR function values will be assigned higher star formation fractions. The proximity of the data point to each sequence influences its final AGN and shock fractions. Depending on the position of the data points on the 3D diagnostic diagram, it is possible that any given data point will be assigned a fraction of 0\% for one or more of the ionising sources.  

\subsubsection{Results}

\begin{table*}
\centering
\begin{tabular}{| c | c | c | c | c | c |}
\toprule
\multicolumn{6}{c}{\textbf{NGC 1068}} \\
\midrule
& \multicolumn{3}{c}{\textbf{New method}} & \multicolumn{2}{c}{\textbf{Old method}} \\
\midrule
\textbf{Emission line} & \textbf{Star formation} & \textbf{Shocks} & \textbf{AGN} & \textbf{Star formation} & \textbf{AGN} \\
\midrule
\textbf{H$\alpha$} & 59.7 $\pm$ 8.0\% & 26.1 $\pm$ 7.4\% & 14.2 $\pm$ 4.5\% & 64.4 $\pm$ 7.1\% & 35.6 $\pm$ 7.1\% \\
\textbf{H$\beta$} & 60.0 $\pm$ 8.1\% & 26.0 $\pm$ 7.4\% & 14.1 $\pm$ 4.5\% & 64.6 $\pm$ 7.1\% & 35.4 $\pm$ 7.1\% \\
\textbf{[O \textsc{ii}]$\lambda \lambda$3726,3729} & 35.4 $\pm$ 10.0\% & 41.2 $\pm$ 12.1\% & 23.3 $\pm$ 8.8\% & 38.3 $\pm$ 12.0\% & 61.7 $\pm$ 12.0\% \\
\textbf{[O \textsc{iii}]$\lambda$5007} & 36.1 $\pm$ 10.8\% & 46.1 $\pm$ 11.5\% & 17.8 $\pm$ 6.5\% & 39.6 $\pm$ 12.8\% & 60.4 $\pm$ 12.8\% \\
\textbf{[S \textsc{ii}]$\lambda \lambda$6716,6731} & 39.4 $\pm$ 9.5\% & 39.1 $\pm$ 11.3\% & 21.5 $\pm$ 8.0\% & 42.5 $\pm$ 11.1\% & 57.5 $\pm$ 11.1\% \\
\textbf{[N \textsc{ii}]$\lambda$6584} & 43.9 $\pm$ 9.5\% & 37.6 $\pm$ 10.3\% & 18.5 $\pm$ 6.7\% & 48.0 $\pm$ 10.4\% & 52.0 $\pm$ 10.4\% \\
\bottomrule
\end{tabular}
\caption{Star formation, shock, and AGN fractions for various strong emission lines in NGC 1068 within the S7 field-of-view. Errors on the star formation, shock, and AGN fractions account for differences in final emission line weighting, uncertainties in photoionisation models, and uncertainities in the fits from \textsc{lzifu}. Displayed also are the star formation and AGN fractions using the method from \citet{TYPHOONpaper}, assuming only star formation-AGN mixing. The errors associated with the star formation-AGN mixing fractions are the result of a 0.1 dex uncertainty in the photoionisation models, and the errors from \textsc{lzifu}.}
\label{tab:agnfracs}
\end{table*}

\begin{figure*}
\centering
\includegraphics[width=\textwidth]{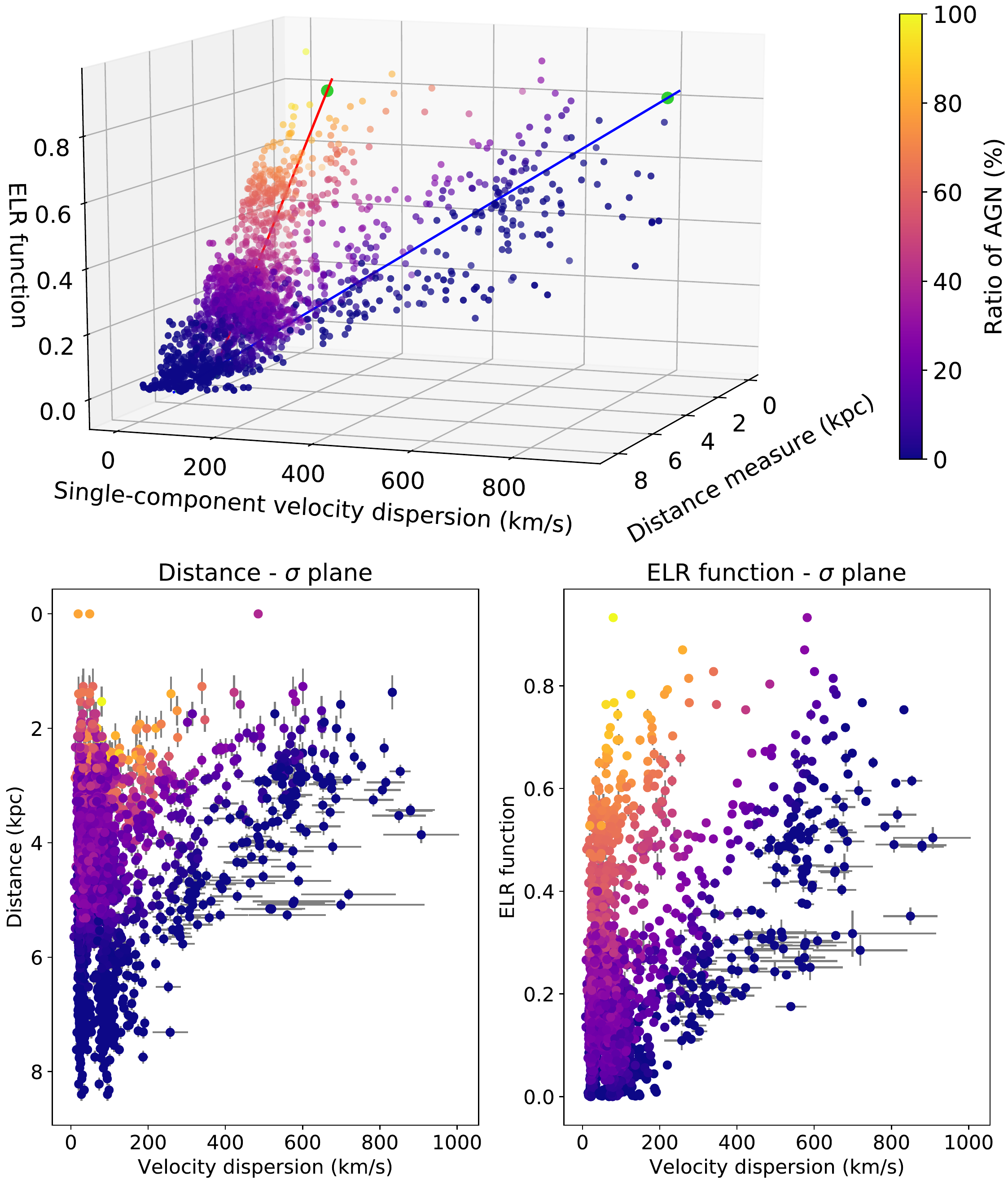}
\caption{3D diagnostic diagram, showing the ratio of AGN in each spaxel of NGC 1068. Light green points represent the basis points for the ELR function. Bottom panels show the 2D projections in the distance-$\sigma$ and ELR function-$\sigma$ planes. Grey lines in the two panels represent the errors associated with each dimension. Errors are omitted from the 3D diagram for clarity.}
\label{fig:3dresults_agn}
\end{figure*}

\begin{figure*}
\centering
\includegraphics[width=\textwidth]{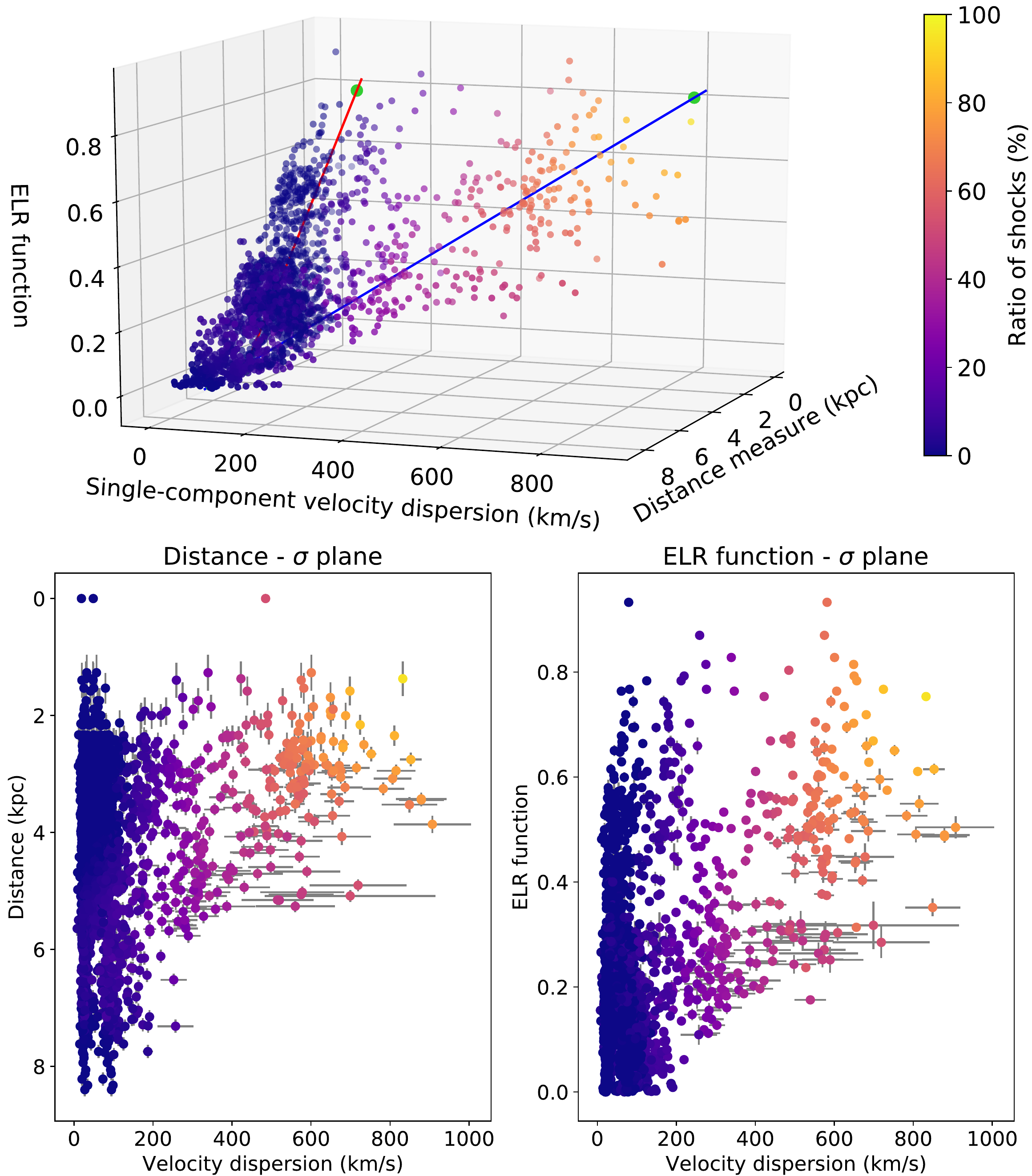}
\caption{3D diagnostic diagram, showing the ratio of shocks in each spaxel of NGC 1068. Light green points represent the basis points for the ELR function. Bottom panels show the 2D projections in the distance-$\sigma$ and ELR function-$\sigma$ planes. Grey lines in the two panels represent the errors associated with each dimension. Errors are omitted from the 3D diagram for clarity.}
\label{fig:3dresults_shocks}
\end{figure*}

\begin{figure*}
\centering
\includegraphics[width=\textwidth]{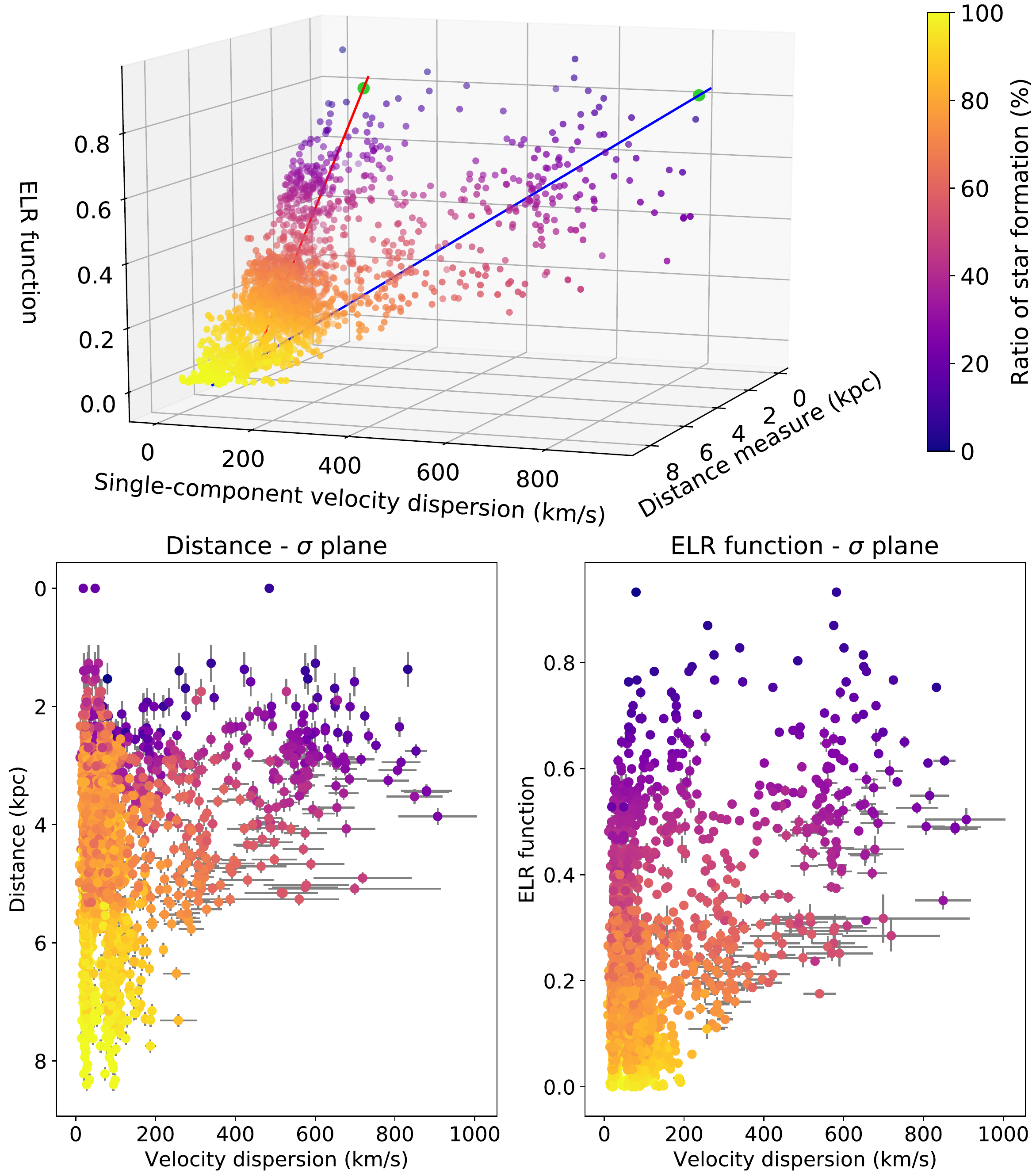}
\caption{3D diagnostic diagram, showing the ratio of star formation in each spaxel of NGC 1068. Light green points represent the basis points for the ELR function. Bottom panels show the 2D projections in the distance-$\sigma$ and ELR function-$\sigma$ planes. Grey lines in the two panels represent the errors associated with each dimension. Errors are omitted from the 3D diagram for clarity.}
\label{fig:3dresults_sf}
\end{figure*}

\begin{figure*}
\centering
\includegraphics[scale=0.55]{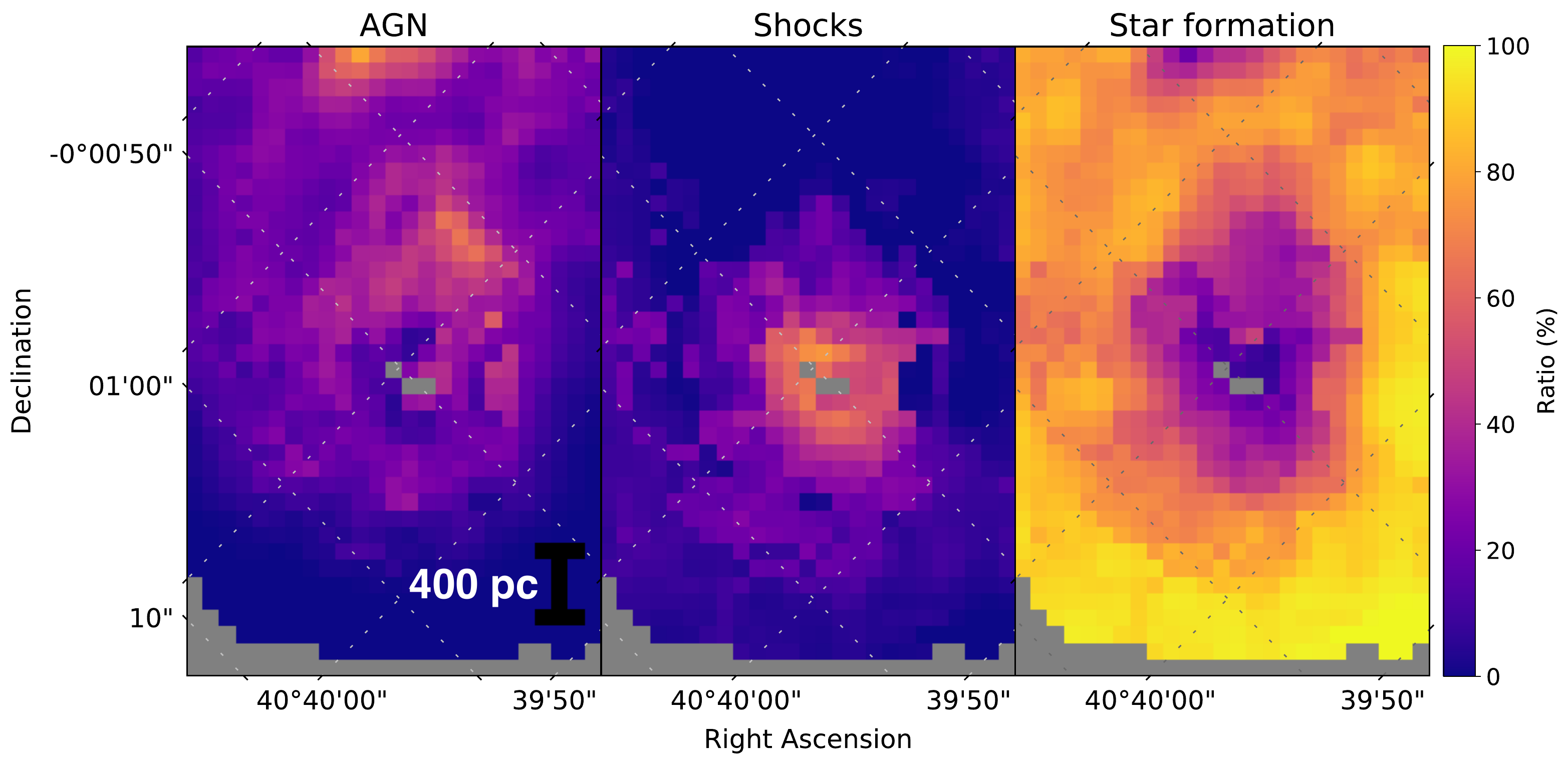}
\caption{Maps of NGC 1068 showing the distribution of the AGN, shock, and star formation fractions seen in Figures~\ref{fig:3dresults_agn}, \ref{fig:3dresults_shocks}, and \ref{fig:3dresults_sf}, respectively. In the event that a spaxel contains more than one velocity component, the final ratio is weighted by the the contribution of each component to the total [O \textsc{iii}]$\lambda 5007$ flux. Dashed lines represent grid lines of constant right ascension and declination.}
\label{fig:mapresults}
\end{figure*}

\begin{figure}
\centering
\begin{subfigure}{\columnwidth}
\includegraphics[width=\textwidth]{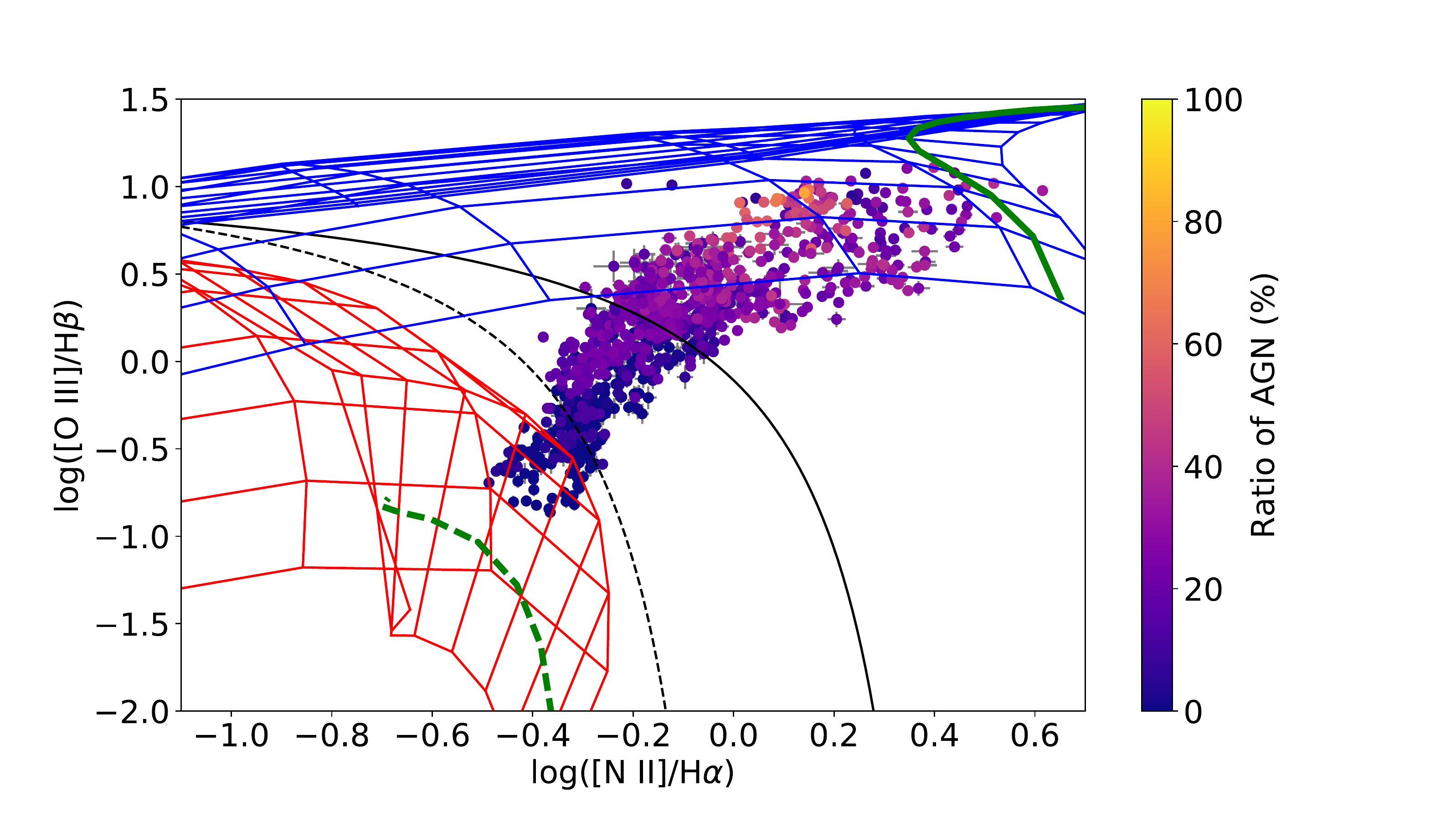}
\label{fig:bptresultsa}
\end{subfigure} 
\begin{subfigure}{\columnwidth}
\includegraphics[width=1\textwidth]{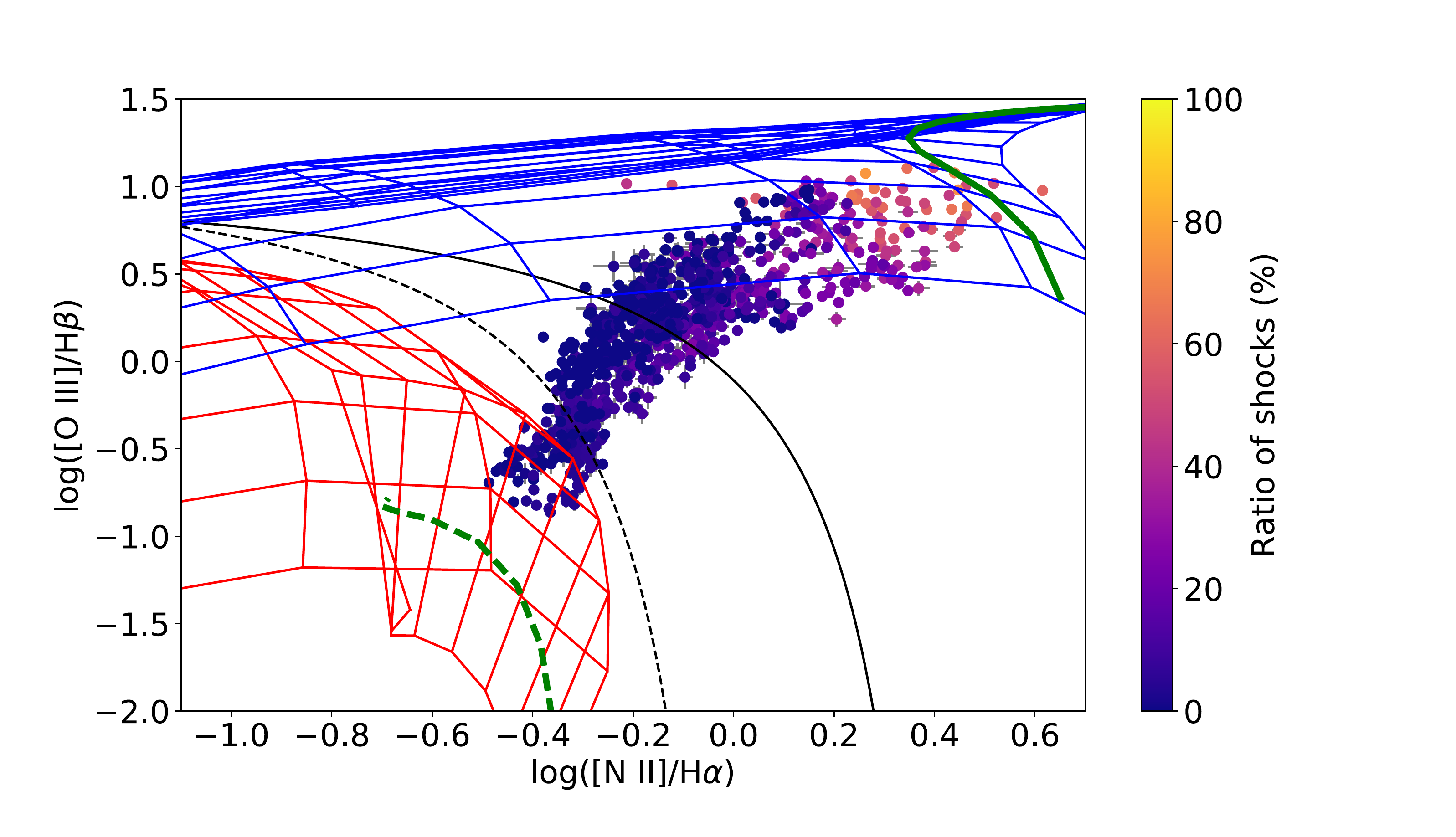}
\label{fig:bptresultsb}
\end{subfigure} 
\begin{subfigure}{\columnwidth}
\includegraphics[width=1\textwidth]{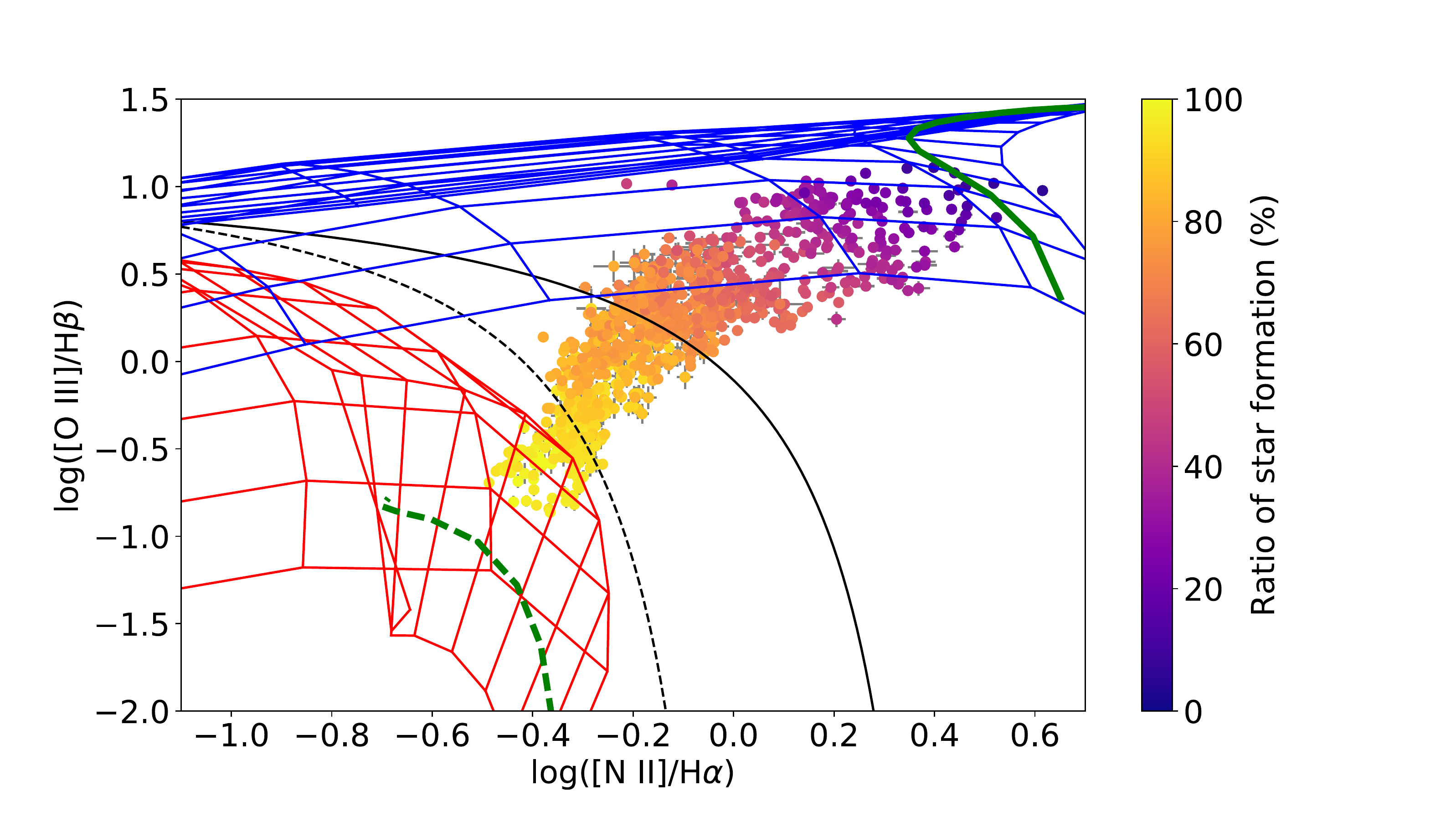}
\label{fig:bptresultsc}
\end{subfigure}
\caption{BPT diagrams showing data from NGC 1068, colour-coded to the AGN, shock, and star formation fractions in each spaxel. The red grid is the H \textsc{ii} region model grid, and the blue grid is the NLR model grid, both described in Section~\ref{sec:models}. The green dashed line on the H \textsc{ii} region grid is a constructed line of constant metallicity for the outer regions of the galaxy, within the S7 field-of-view. The solid green line on the NLR model is the constructed line of constant metallicity for the centre of the galaxy. Both metallicity values are found in Table~\ref{tab:values}. The dashed black line is the \citet{Kauffmann2003} demarcation line, and the solid black line is the \citet{Kewley2001} demarcation line.}
\label{fig:bptresults}
\end{figure}

The results of calculating AGN, shock, and star formation fractions on the 3D diagnostic diagram for NGC 1068 are shown in Figures~\ref{fig:3dresults_agn}, \ref{fig:3dresults_shocks}, and \ref{fig:3dresults_sf}, respectively. We also include 2D projections of the fractions in the distance-$\sigma$ and ELR function-$\sigma$ planes in Figures~\ref{fig:3dresults_agn}, \ref{fig:3dresults_shocks}, and \ref{fig:3dresults_sf}. Unlike in Figure~\ref{fig:3ddiags}, we do not include the ELR function-distance plane projection, due to the obscuration of many spaxels preventing its usefulness. A line of best fit is applied to both the star formation-AGN and star formation-shock sequences of spaxels to show the direction of star formation-AGN and star formation-shock mixing respectively. The fractions are calculated for all spaxels on the 3D diagnostic diagram; some spaxels may at this point have multiple values for their star formation-shock-AGN fractions, depending on the number of velocity components present in the spaxel. As a result, for a spaxel with multiple velocity components, the final fractions are calculated as the weighted average of the fractions in the individual components, weighted by the contribution of each component to the total [O \textsc{iii}]$\lambda 5007$ flux. Maps of the final star formation, shock, and AGN fractions across the galaxy in the S7 field-of-view are shown in Figure~\ref{fig:mapresults}. BPT diagrams of NGC 1068, with each spaxel colour-coded to the final star formation, shock, and AGN fractions, are shown in Figure~\ref{fig:bptresults}. We have chosen [O \textsc{iii}]$\lambda 5007$ as the line by which we calculate the relative weight of each component, because it is a strong emission line associated with the presence of an AGN and shocks. Hence, it is a reliable emission line to use for the analysis in this work. However, other strong lines are present in the data, such as H$\alpha$ and [N \textsc{ii}]$\lambda 6584$. We calculate that the choice of which emission line is used to calculate the weighting can impact the final fractions by up to ${\sim} 8$\%. It is important to note that if two galaxies are to be compared equally, the emission line by which the final fractions are weighted must be identical. When calculating the relative contribution of each ionising source to the flux of strong emission lines in Table~\ref{tab:agnfracs} (described below), the errors on the fractions account for the possible differences in value as a result of weighting by a different emission line.

The star-forming ring in NGC 1068 can be seen surrounding the nucleus in the star formation ratio map from Figure~\ref{fig:mapresults}. The star formation ratio map also shows that ionisation in the nucleus is dominated by non-stellar sources. In particular, the biconical outflow most prominently seen in the [O \textsc{iii}]$\lambda$5007 is largely visible in the AGN ratio map. This agrees with previous work, claiming that the bicone is the result of material radiatively accelerated by the AGN \citep[e.g.][]{Pogge1988,Cecil2002,Dopita2002b,TYPHOONpaper}. The shock ratio map shows that shocks are seen to dominate the central few arcseconds of NGC 1068, however, due to the location of the high shock fractions, these shocks are likely from the AGN. An accreting supermassive black hole will produce thermal X-rays which provide a radiation pressure on nearby gas, causing expulsion of the gas in the form of a wind. This high-velocity wind will cause shocked material upon interaction with the ISM \citep{ZK2012}. Hence, the shocked structure seen in the nucleus of NGC 1068 is likely the result of the AGN.


The star formation, shock, and AGN contributions to the flux of several strong lines are shown in Table~\ref{tab:agnfracs}. For comparison, also in Table~\ref{tab:agnfracs}, we show the star formation and AGN contributions to the same strong emission lines, calculated using the star formation-AGN mixing method from \citet{TYPHOONpaper}. The method used to calculate the star formation, shock, and AGN contributions to each emission line listed in Table~\ref{tab:agnfracs} is identical to the method described in \citet{Davies2014a} and \citet{TYPHOONpaper}:

The total luminosity of any emission line in an IFU field-of-view is given by:
\[ L_{\mathrm{Tot}} = \sum_{i = 1}^{n} L_i \]
where $L_i$ is the luminosity of the emission line in spaxel $i$. The total luminosity of the emission line attributable to any source (star formation, shocks, or AGN) can be calculated by
\[ L_{\mathrm{source}} = \sum_{i = 1}^{n} f_{i}^{\mathrm{source}} L_i \]
where $f_{i}^{\mathrm{source}}$ is the source fraction in spaxel $i$. It follows that the relative fraction of emission attributable to each source for the given emission line is provided by
\[ f_{\mathrm{Tot}}^{\mathrm{source}} = \frac{L_{\mathrm{source}}}{L_{\mathrm{Tot}}} \] 

In addition to accounting for the choice of emission line used for the final weightings, the errors associated with each fraction given in Table~\ref{tab:agnfracs} also consider the error in the photoionisation models described in Section~\ref{sec:models}. Assuming an error of ${\sim} 0.1$ dex in both the [O \textsc{iii}]/H$\beta$ and [N \textsc{ii}]/H$\alpha$ ratios \citep[e.g.][]{Kewley2001,gridpaper}, the resulting change in the position of the basis points on the 3D diagram can alter the final fractions by up to ${\sim} 10$\%. Finally, the errors on the fractions in Table~\ref{tab:agnfracs} also consider the uncertainties in the emission line fits from \textsc{lzifu}. The error analysis performed on the star formation-AGN mixing fractions also in Table~\ref{tab:agnfracs} is updated from that in \citet{TYPHOONpaper}, considering the 0.1 dex uncertainty in the photoionisation models.

\begin{figure*}
\centering
\includegraphics[scale=0.53]{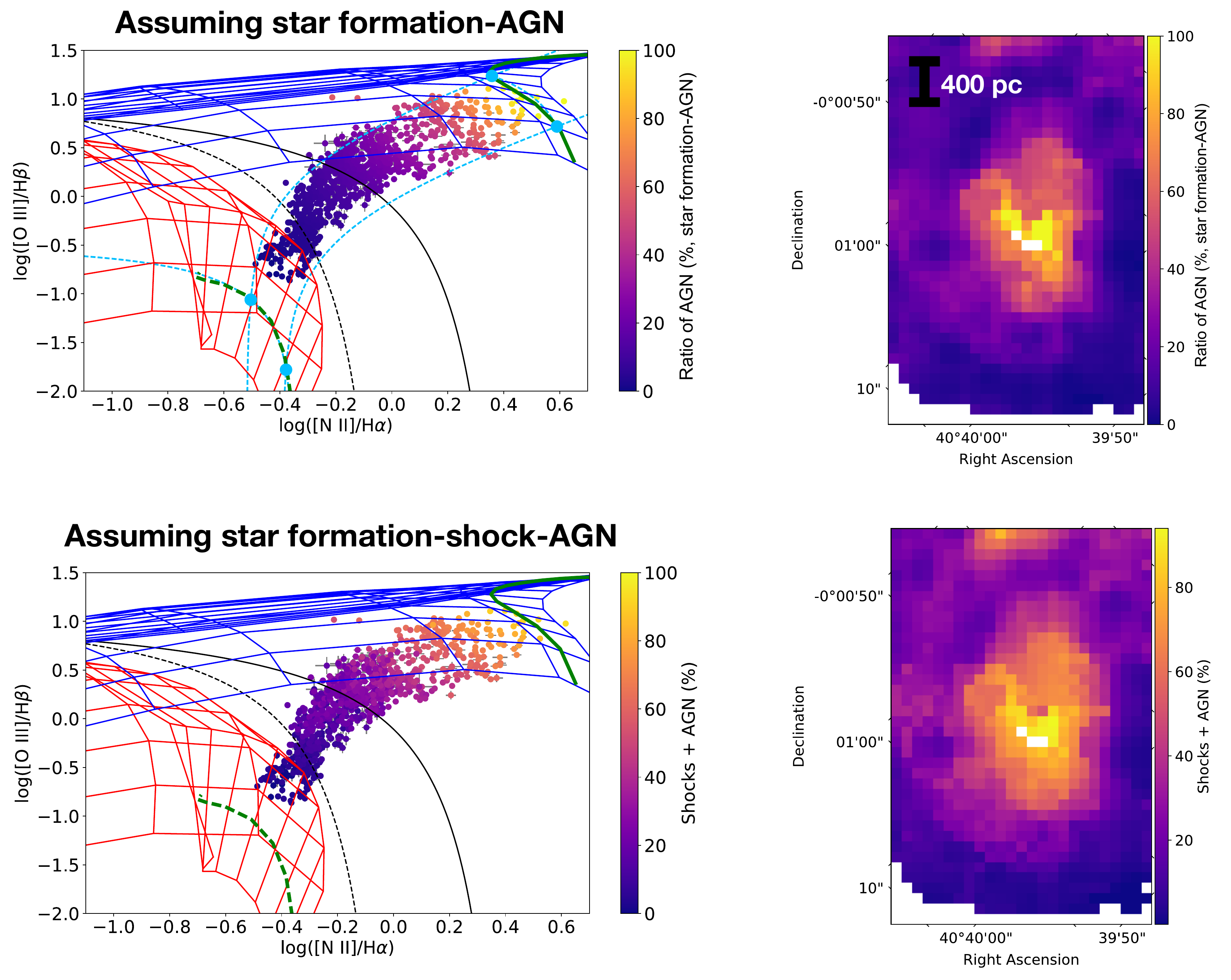}
\caption{Maps and BPT diagrams of NGC 1068, assuming a star formation-AGN mixing regime on top, and our new star formation-shock-AGN regime below. The spaxels in the bottom BPT diagram and map are coloured by shock + AGN fraction, seen separately in Figures~\ref{fig:mapresults} and \ref{fig:bptresults}. The red grid on the BPT diagrams is the H \textsc{ii} region model grid, and the blue grid is the NLR model grid, both described in Section~\ref{sec:models}. The green dashed line on the H \textsc{ii} region grid is a constructed line of constant metallicity for the outer regions of the galaxy, within the S7 field-of-view. The solid green line on the NLR model is the constructed line of constant metallicity for the centre of the galaxy. Both metallicity values are found in Table~\ref{tab:values}. The light blue points are the basis points used to calculate the AGN fraction in each spaxel \citep[see][]{TYPHOONpaper}. The dashed black line is the \citet{Kauffmann2003} demarcation line, and the solid black line is the \citet{Kewley2001} demarcation line.}
\label{fig:sandp}
\end{figure*}

The method from \citet{TYPHOONpaper} assumes the emission in each spaxel can be expressed as a combination of star formation and AGN activity, without accounting for shock emission. The fractions in Table~\ref{tab:agnfracs} show that if shocks are not considered as a possible source of ionisation, then the contribution to photoionisation from an AGN may be severely overestimated. The star formation fractions in each of the strong lines listed are similar between the two methods, implying that the majority of shock emission mixes with the emission from AGN photoionisation. This is further illustrated in Figure~\ref{fig:sandp}. We show a BPT diagram and map of NGC 1068 coloured by AGN fraction, assuming only a star formation-AGN mixing regime, against a BPT diagram and map coloured by the shock $+$ AGN fraction, calculated by adding the fractions seen in Figures~\ref{fig:mapresults} and~\ref{fig:bptresults}. The results are very similar. This is unsurprising, given that shocks can produce emission line ratios similar to that of AGN. This is particularly true on the BPT diagram, where emission from shocks and AGN can be found within the same region of the diagram. Hence, using the BPT diagram to calculate star formation and AGN fractions, and only assuming a star formation-AGN mixing regime as in \citet{Davies2014a,Davies2014b}, \citet{Davies2016}, and \citet{TYPHOONpaper}, will likely overestimate the true contribution from the AGN to emission line fluxes.

\citet{Davies2014a,Davies2014b}, \citet{Davies2016}, and \citet{TYPHOONpaper} all show that using H$\alpha$ as a SFR indicator in an AGN galaxy may lead to an overestimate of the true SFR, as the AGN is responsible for a fraction of the total H$\alpha$ luminosity. This is also seen in Table~\ref{tab:agnfracs}, as the AGN contribution to the H$\alpha$ flux in NGC 1068 is non-zero, irrespective of the inclusion of shocks to the decomposition. However, once shocks are accounted for in the decomposition, the star formation fractions of each of the strong lines, while similar, are systematically lower. Therefore, it is evident that emission from star formation does not solely mix with AGN, but rather multiple sources of ionisation and excitation. In order to calculate a SFR as accurately as possible using the H$\alpha$ flux, as many sources of ionisation and excitation as possible must be simultaneously separated. Failure to do so will result in the SFR consistently being overestimated. Using the SFR(H$\alpha$) relation given by \citet{Kennicutt1994}, we calculate a SFR of 3.1 $M_\odot$ yr$^{-1}$ in the S7 field-of-view (radius ${\sim} 1$ kpc), after calculating the H$\alpha$ luminosity from star formation to be $3.9 \times 10^{41}$ erg s$^{-1}$.

\section{Comparison with other data}
\label{sec:wavelengths}

To verify our star formation, shock, and AGN fractions in NGC 1068, we use the structure seen in other wavelengths. X-ray emission is known to be ubiquitous in AGN \citep[e.g.][]{Gandhi2005}, and nuclear shocks may be caused by the relativistic jets from the SMBH, which are very visible in the radio part of the spectrum \citep{BK1979}. We also study the CO(3-2) distribution in the nucleus of NGC 1068 to verify our star formation results. \citet{GB2014} show that the majority of CO(3-2) flux measured by ALMA (${\sim} 63$\% of the total) is detected in the star-forming ring surrounding the nucleus.

\begin{figure*}
\centering
\includegraphics[scale=0.6]{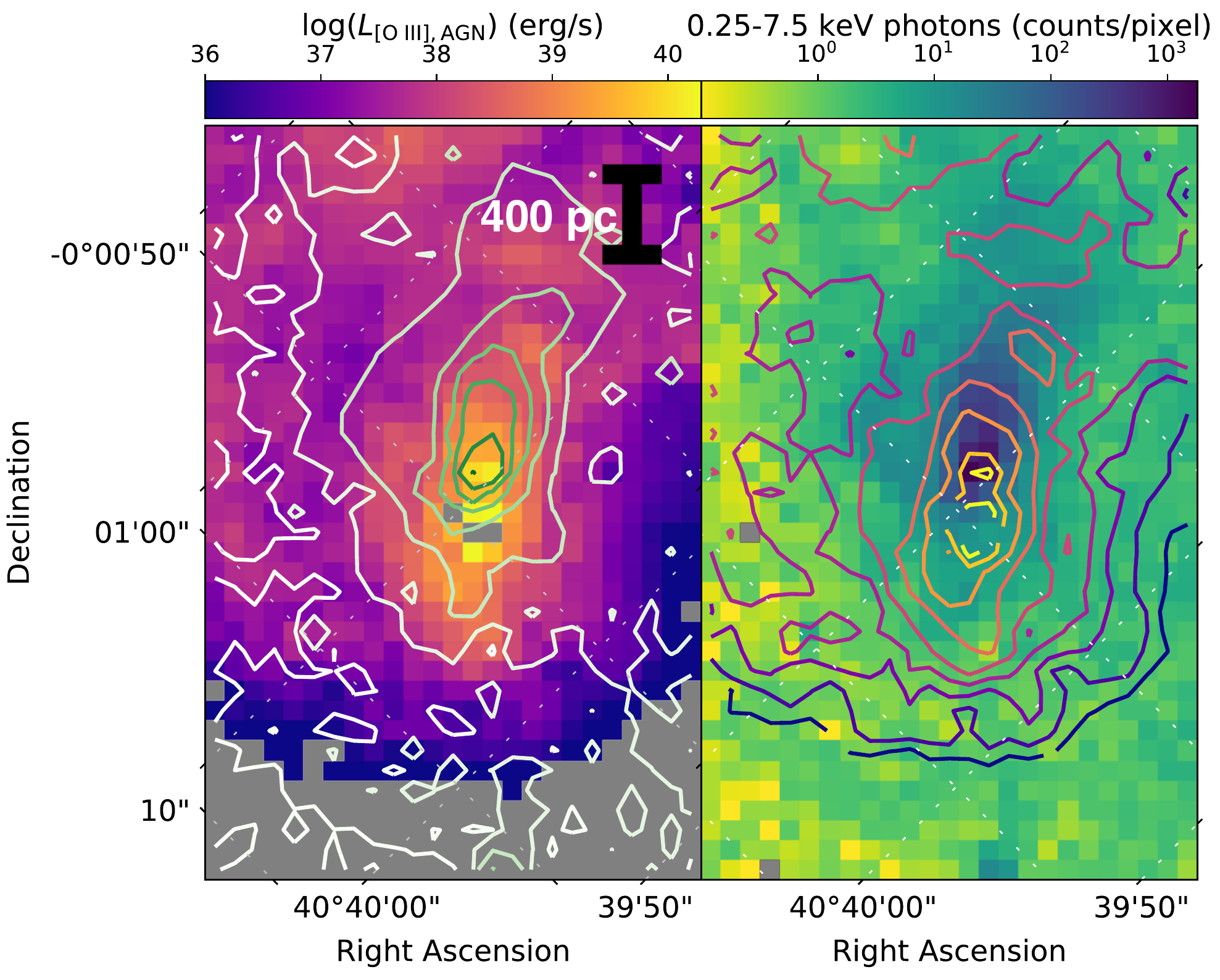}
\caption{Maps of NGC 1068 showing the [O \textsc{iii}]$\lambda5007$ luminosity attributable to AGN in the left panel, and the 0.25-7.5 keV X-ray photon map from \citet{Young2001} on the right. Contours of each map are shown in the adjacent panel. Dashed lines represent grid lines of constant right ascension and declination.}
\label{fig:xraycont}
\end{figure*}

\begin{figure*}
\centering
\includegraphics[scale=0.6]{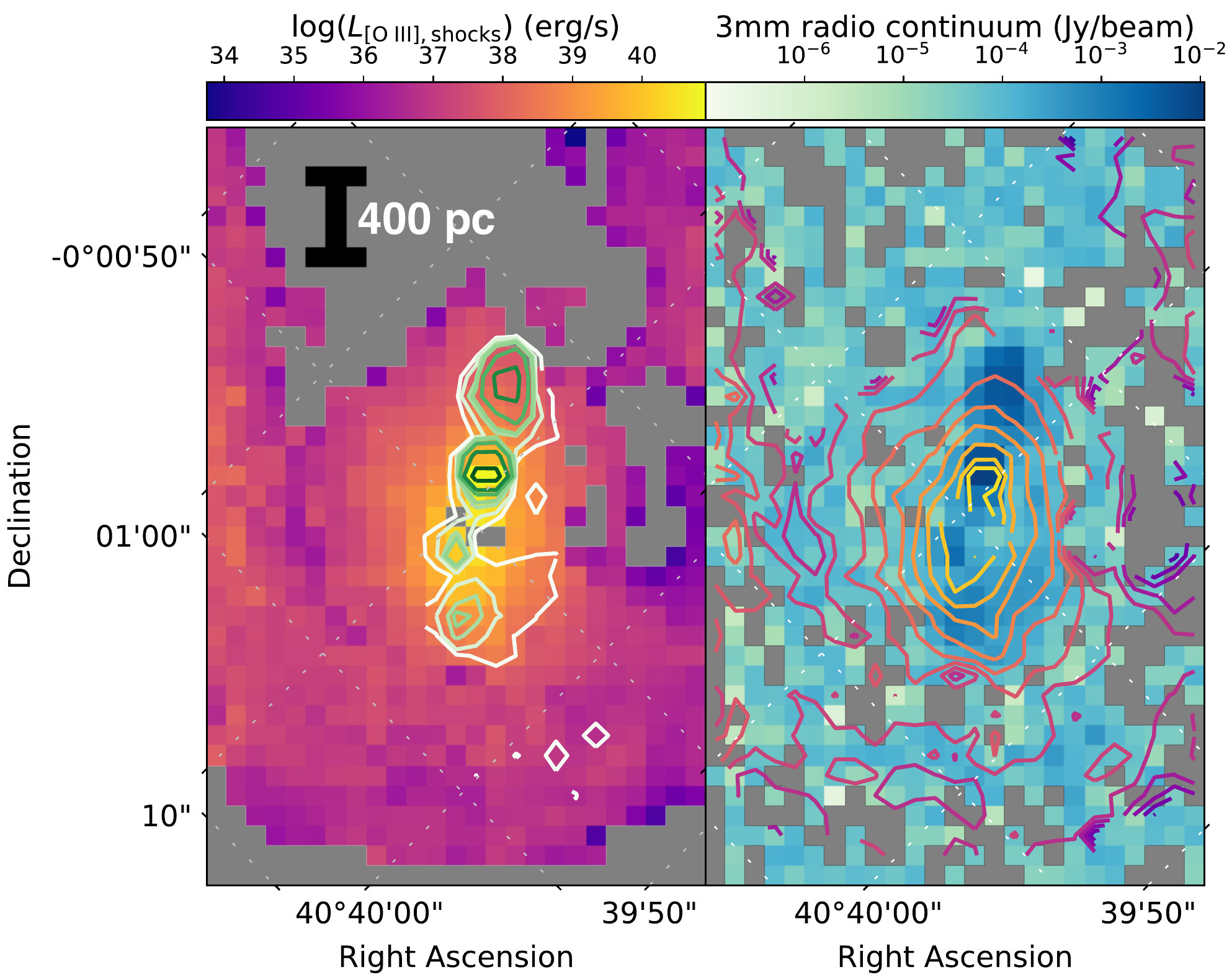}
\caption{Maps of NGC 1068 showing the [O \textsc{iii}]$\lambda5007$ luminosity attributable to shocks in the left panel, and the 3mm radio continuum map from \citet{GB2017} on the right. Contours of each map are shown in the adjacent panel. Dashed lines represent grid lines of constant right ascension and declination.}
\label{fig:radiocont}
\end{figure*}

\begin{figure*}
\centering
\includegraphics[scale=0.6]{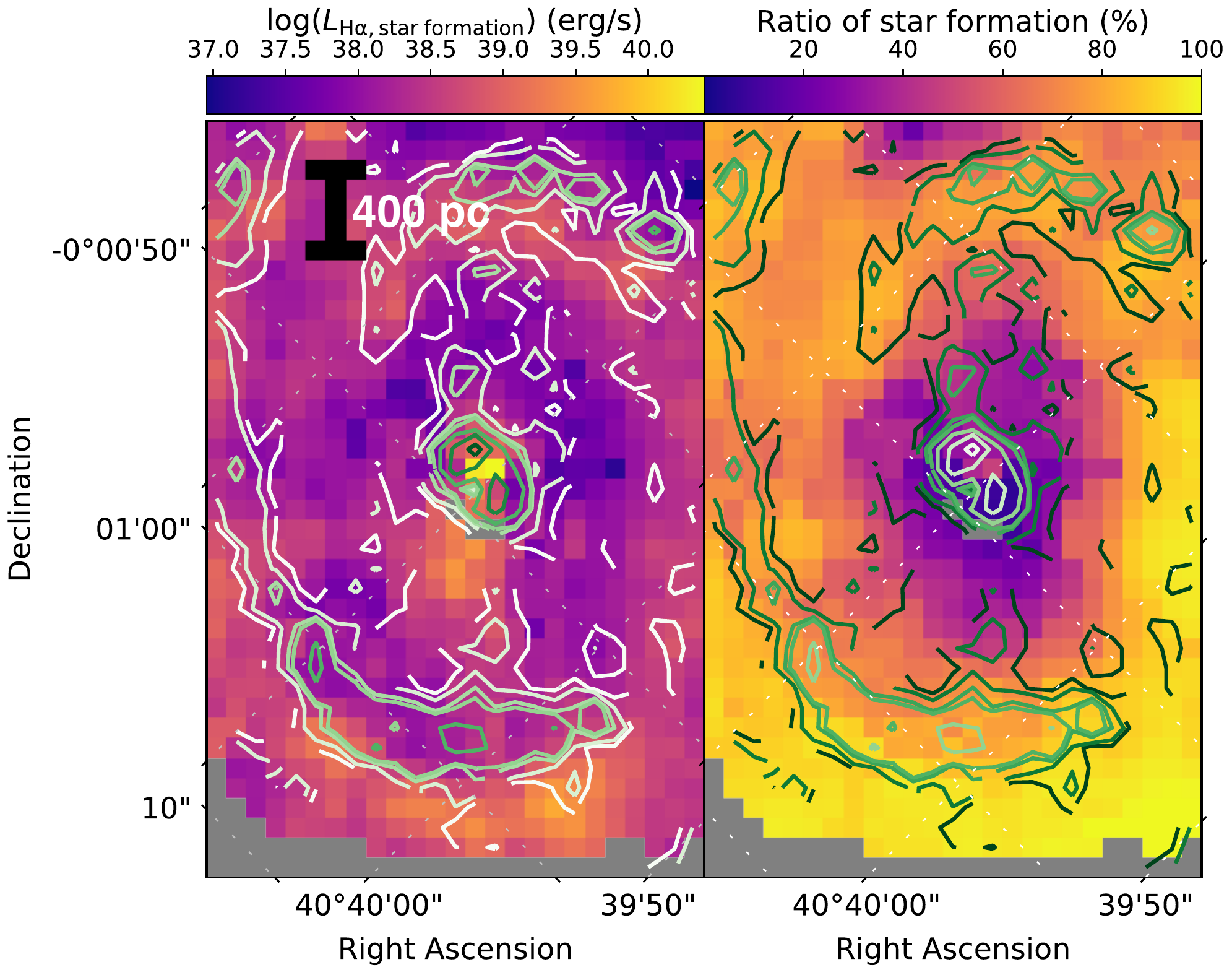}
\caption{Maps of NGC 1068 showing the H$\alpha$ luminosity attributable to star formation in the left panel, and the star formation ratio map from Figure~\ref{fig:mapresults} on the right. Identical contours of the CO(3-2) flux from \citet{GB2014} are shown on both maps. Dashed lines represent grid lines of constant right ascension and declination.}
\label{fig:sbcont}
\end{figure*}

Shown in Figure~\ref{fig:xraycont} are maps of the [O \textsc{iii}]$\lambda5007$ luminosity from the AGN, and the 0.25-7.5 keV \textit{Chandra} X-ray distribution from \citet{Young2001}. Contours of each map are shown on the other. The X-ray contours accurately trace the regions of high luminosity in the AGN map. Similarly, we find large agreement with the distribution of the [O \textsc{iii}]$\lambda5007$ shock luminosity, and the 3mm radio continuum contours from \citet{GB2017} in Figure~\ref{fig:radiocont}. The 3mm radio continuum data from \citet{GB2017} clearly shows the distribution and position of the radio jets. Finally, in Figure~\ref{fig:sbcont} we show the CO(3-2) contours from \citet{GB2014} over the H$\alpha$ luminosity distribution from star formation, and the star formation ratio map from Figure~\ref{fig:mapresults}. The CO(3-2) contours accurately trace regions of high star formation surrounding the nucleus, in agreement with the findings of \citet{GB2014}. 

\begin{figure*}
\centering
\begin{minipage}{0.45\textwidth}
\centering
\includegraphics[scale=0.54]{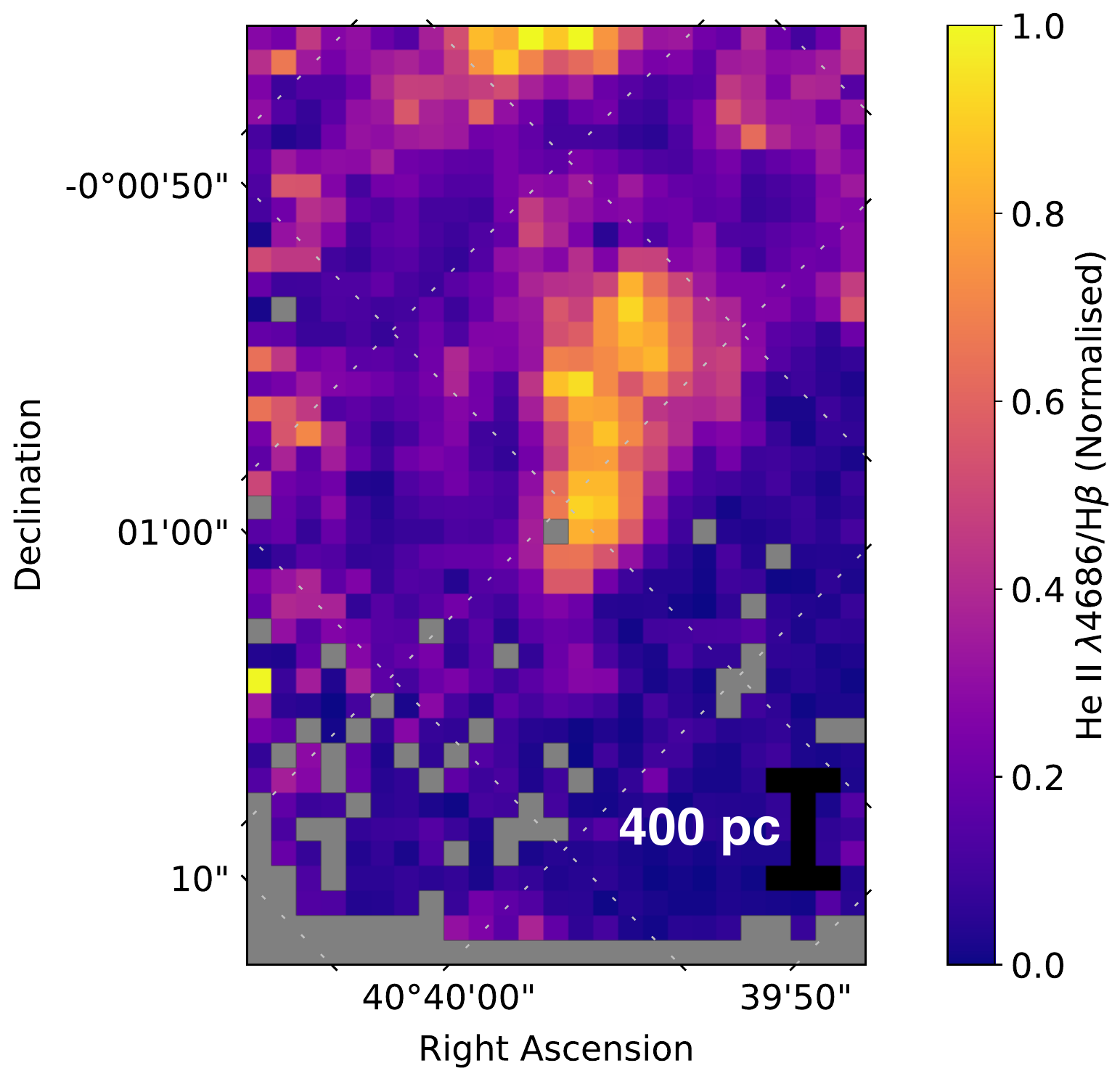}
\caption{He \textsc{ii}/H$\beta$ distribution map for NGC 1068. The He \textsc{ii}/H$\beta$ ratio in each spaxel has been normalised to the maximum value (He \textsc{ii}/H$\beta_\mathrm{max} \sim 0.49$). The He \textsc{ii}/H$\beta$ emission is seen to form a clean one-sided cone. Dashed lines represent grid lines of constant right ascension and declination.}
\label{fig:heiihb}
\end{minipage}\hfill
\begin{minipage}{0.45\textwidth}
\centering
\includegraphics[scale=0.37]{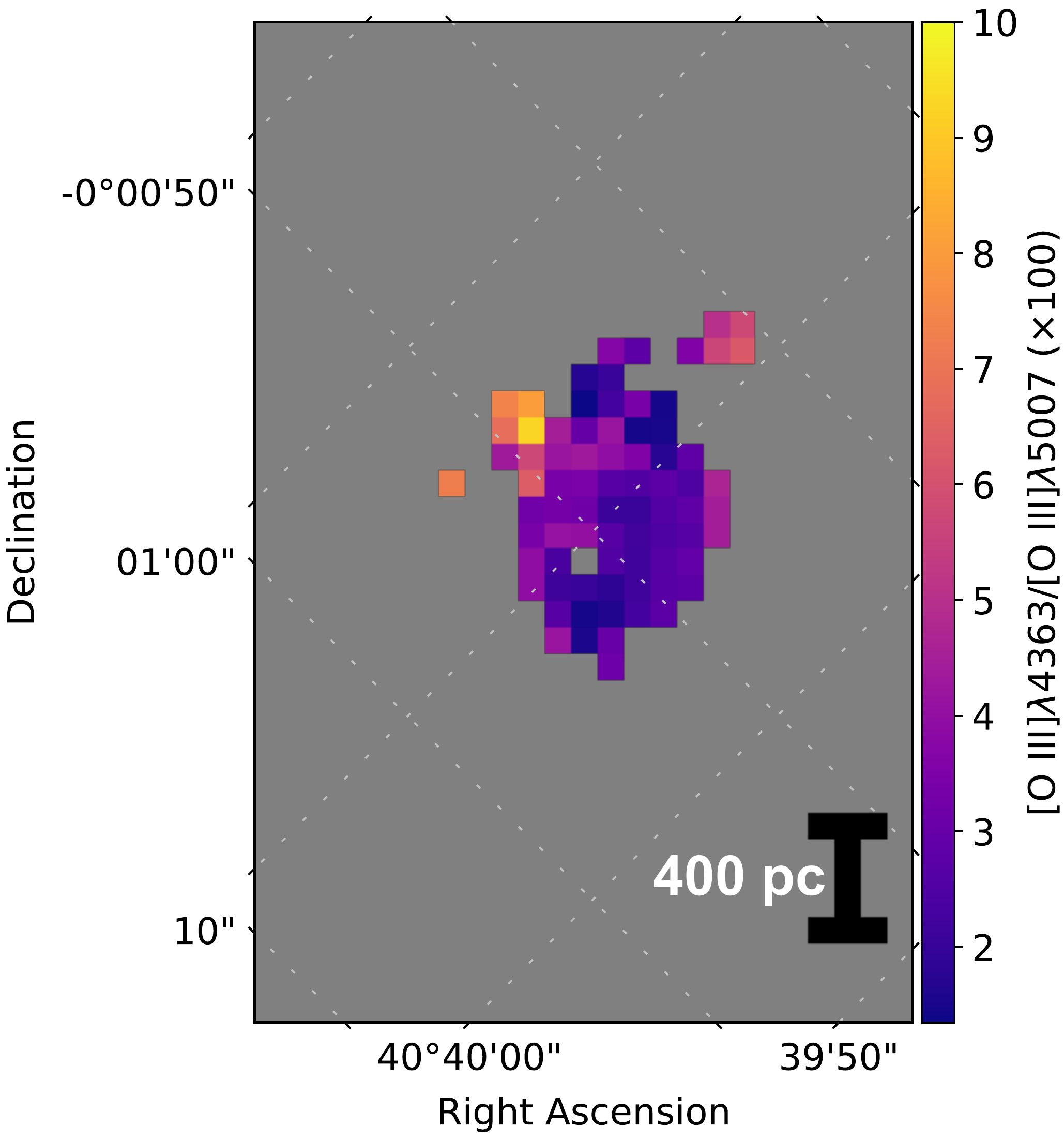}
\caption{Distribution of the temperature-sensitive [O \textsc{iii}]$\lambda 4363$/[O \textsc{iii}]$\lambda 5007$ line ratio in NGC 1068. Only spaxels with significant detection of both the [O \textsc{iii}]$\lambda 4363$ and [O \textsc{iii}]$\lambda 5007$ emission lines are shown (signal-to-noise $> 3$). Significant spaxels are seen to align with spaxels containing high shock + AGN fractions in Figure~\ref{fig:mapresults}. Dashed lines represent grid lines of constant right ascension and declination.}
\label{fig:oiiiratio}
\end{minipage}
\end{figure*}



We also use of the He \textsc{ii}$\lambda 4686$/H$\beta$ distribution map to verify our results. He \textsc{ii} is a high-ionisation line (ionisation potential of ${\sim} 54$ eV), thus requiring a hard radiation field (the kind associated with AGN) for ionisation. The line ratio He \textsc{ii}/H$\beta$ is not contained in our emission line ratio function shown in Equation~\ref{eq:elr}, and so provides an independent test of our results. The He \textsc{ii}/H$\beta$ map from the S7 data is shown in Figure~\ref{fig:heiihb}, normalised to the maximum value. Regions of high He \textsc{ii}/H$\beta$ can be seen to align with regions of high AGN fraction in Figure~\ref{fig:mapresults}, as expected. 

To further verify our shock results, we consider the distribution of the [O \textsc{iii}]$\lambda 4363$/[O \textsc{iii}]$\lambda 5007$ ratio across the galaxy. The [O \textsc{iii}] ratio is a temperature-sensitive line ratio, and increases in the line ratio indicate an increase in the electron temperature. Shocked gas can reach temperatures on the order of $10^6$ K \citep[for a fully-ionised plasma; e.g.][]{ADU}, compared to the NLR [O \textsc{iii}] temperatures of ${\sim} 20,000$ K discussed in Section~\ref{sec:models}. Hence, an increase in the [O \textsc{iii}] line ratio is expected to coincide with regions of high shock fractions in Figure~\ref{fig:mapresults}.

The [O \textsc{iii}] line ratio map is shown in Figure~\ref{fig:oiiiratio}. Only spaxels which contain significant detection of both the [O \textsc{iii}]$\lambda 4363$ and [O \textsc{iii}]$\lambda 5007$ emission lines are shown in Figure~\ref{fig:oiiiratio}, corresponding to a signal-to-noise greater than 3. Such spaxels are concentrated towards the centre, where the [O \textsc{iii}] luminosity is high, seen for example in Figures~\ref{fig:xraycont} and~\ref{fig:radiocont}. The variation in the [O \textsc{iii}] ratio amongst included spaxels is low, and hence the values in Figure~\ref{fig:oiiiratio} have been enhanced $\times 100$ to better highlight the variation. Clearly, an increase in the [O \textsc{iii}] ratio is seen towards the top-left of the included spaxels. Similarly, the shock fraction map in Figure~\ref{fig:mapresults} also shows a slight increase in the shock fractions in spaxels towards the top-left of centre. Although, it should be noted that in this region, the AGN fractions of the spaxels also seen in Figure~\ref{fig:mapresults} are considerably high (${\sim} 40$\%). Furthermore, photoionisation modelling with \textsc{mappings} shows that AGN photoionisation can lead to an increase in the [O \textsc{iii}] ratio, in the range from that expected from fast shocks of ${\sim} 0.1 - 0.01$ \citep[e.g.][]{Dopita2002,Allen2008}. The variation seen in the [O \textsc{iii}] ratio map is better aligned with the features in the shock + AGN map in Figure~\ref{fig:sandp}. Thus, we conclude that the distribution of the [O \textsc{iii}] ratio does not show sufficient variation to make confident assumptions about the shock-to-AGN ratio.

The shock-dominated regions are primarily located close to the nucleus of NGC 1068. Thus, possible mechanisms for their creation include NLR outflows, or the associated radio structures. We suggest that the shocks are the result of NLR outflows. The radio jet structure seen in \textit{HST} imaging from \citet{Cecil2002} shows a radio structure at the nucleus of NGC 1068 with an extent of ${\sim} 1 - 2$ arcseconds. This is far smaller than the extent of the shock-dominated spaxels, which can extend beyond 10 arcseconds, seen in Figure~\ref{fig:mapresults}. In addition, our Figure~\ref{fig:vdispmaps}, and Figure 4 from \citet{3dletter} show the extent of the high-$\sigma$ spaxels to be far beyond ${\sim} 1 - 2$ arcseconds. Hence, the small-scale radio jet must not be the dominant shock-production mechanism. 


\section{Conclusions and future work}
\label{sec:conclusions}

We have demonstrated a new method to simultaneously separate line emission from gas ionised by star formation, shocks, and AGN in IFU data. Our 3D diagnostic diagram shows two clear mixing sequences, indicating star formation-AGN and star formation-shock mixing within a single galaxy. Using the information in these two clear mixing sequences, we have quantified the ratio of star formation-, shock-, and AGN-ionised line emission in each spaxel of the Seyfert galaxy NGC 1068.

We have shown that if shocks are not considered as an ionising source, then the relative contribution from an AGN to emission line fluxes may be greatly overestimated. The shock + AGN fractions for each emission line are very similar to the AGN fractions calculated assuming only a star formation-AGN mixing regime, using the method from \citet{TYPHOONpaper}. This indicates that shocked line emission mixes heavily with the AGN-ionised line emission. Including shocks as a possible ionising source also systematically lowers the star-forming fractions calculated in each emission line. If the flux of H$\alpha$ is to be used as a SFR indicator, it is recommended to separate as many ionsing sources as possible.

We have compared our results to data in various wavelengths. These comparisons are summarised:

(i) The luminosity distribution of AGN-ionised line emission closely resembles the 0.25-7.5 keV X-ray photon map from \citet{Young2001}.

(ii) The luminosity distribution of shock-ionised line emission is accurately aligned with the radio jets in NGC 1068, seen in the 3mm radio continuum map from \citet{GB2017}.

(iii) The luminosity distribution of star formation-ionised line emission closely traces the CO(3-2) molecular line flux from \citet{GB2014}. The star-forming ring in NGC 1068 is host to the majority (${\sim} 63$\%) of the CO(3-2) flux in the ALMA field-of-view \citep{GB2014}.

(iv) Regions of AGN-dominated line emission correlate very accurately with regions of high He \textsc{ii}/H$\beta$ ratios. He \textsc{ii}$\lambda 4686$ is a high-ionisation line, and thus requires a hard radiation field (such as that from an AGN) to produce large emission line fluxes.

(v) Regions of shock-dominated line emission show a slight correlation with regions of increased [O \textsc{iii}]$\lambda 4363$/[O \textsc{iii}]$\lambda 5007$, although the AGN contribution in these regions is also considerable. This indicates that the shock and AGN separation may need further advancements.

In a future publication, we aim to use our star formation-shock-AGN separation method to quantify the star formation, shock, and AGN fractions of merging galaxies, at various merger stages.

\section*{Acknowledgements}

Parts of this research were conducted by the Australian Research Council Centre of Excellence for All Sky Astrophysics in 3 Dimensions (ASTRO 3D), through project number CE170100013. 

Support for AMM is provided by NASA through Hubble Fellowship grant \#HST-HF2-51377 awarded by the Space Telescope Science Institute, which is operated by the Association of Universities for Research in Astronomy, Inc., for NASA, under contract NAS5-26555.    

The authors wish to also acknowledge the contribution from the reviewer, who facilitated great discussion amongst the co-authors and thus greatly improved the paper.



\bibliographystyle{mnras}
\bibliography{ssagn} 

\begin{thebibliography}{}
\makeatletter
\relax
\def\mn@urlcharsother{\let\do\@makeother \do\$\do\&\do\#\do\^\do\_\do\%\do\~}
\def\mn@doi{\begingroup\mn@urlcharsother \@ifnextchar [ {\mn@doi@}
  {\mn@doi@[]}}
\def\mn@doi@[#1]#2{\def\@tempa{#1}\ifx\@tempa\@empty \href
  {http://dx.doi.org/#2} {doi:#2}\else \href {http://dx.doi.org/#2} {#1}\fi
  \endgroup}
\def\mn@eprint#1#2{\mn@eprint@#1:#2::\@nil}
\def\mn@eprint@arXiv#1{\href {http://arxiv.org/abs/#1} {{\tt arXiv:#1}}}
\def\mn@eprint@dblp#1{\href {http://dblp.uni-trier.de/rec/bibtex/#1.xml}
  {dblp:#1}}
\def\mn@eprint@#1:#2:#3:#4\@nil{\def\@tempa {#1}\def\@tempb {#2}\def\@tempc
  {#3}\ifx \@tempc \@empty \let \@tempc \@tempb \let \@tempb \@tempa \fi \ifx
  \@tempb \@empty \def\@tempb {arXiv}\fi \@ifundefined
  {mn@eprint@\@tempb}{\@tempb:\@tempc}{\expandafter \expandafter \csname
  mn@eprint@\@tempb\endcsname \expandafter{\@tempc}}}

\bibitem[\protect\citeauthoryear{{Alexander} \& {Hickox}}{{Alexander} \&
  {Hickox}}{2012}]{AH2012}
{Alexander} D.~M.,  {Hickox} R.~C.,  2012, \mn@doi [\nar]
  {10.1016/j.newar.2011.11.003}, \href
  {http://adsabs.harvard.edu/abs/2012NewAR..56...93A} {56, 93}

\bibitem[\protect\citeauthoryear{{Allen}, {Groves}, {Dopita}, {Sutherland}  \&
  {Kewley}}{{Allen} et~al.}{2008}]{Allen2008}
{Allen} M.~G.,  {Groves} B.~A.,  {Dopita} M.~A.,  {Sutherland} R.~S.,
  {Kewley} L.~J.,  2008, \mn@doi [\apjs] {10.1086/589652}, \href
  {https://ui.adsabs.harvard.edu/abs/2008ApJS..178...20A} {178, 20}

\bibitem[\protect\citeauthoryear{{Anders} \& {Grevesse}}{{Anders} \&
  {Grevesse}}{1989}]{AG1989}
{Anders} E.,  {Grevesse} N.,  1989, \mn@doi [\gca]
  {10.1016/0016-7037(89)90286-X}, \href
  {http://adsabs.harvard.edu/abs/1989GeCoA..53..197A} {53, 197}

\bibitem[\protect\citeauthoryear{{Antonucci} \& {Miller}}{{Antonucci} \&
  {Miller}}{1985}]{AM1985}
{Antonucci} R.~R.~J.,  {Miller} J.~S.,  1985, \mn@doi [\apj] {10.1086/163559},
  \href {http://adsabs.harvard.edu/abs/1985ApJ...297..621A} {297, 621}

\bibitem[\protect\citeauthoryear{{Baldwin}, {Phillips}  \&
  {Terlevich}}{{Baldwin} et~al.}{1981}]{BPT1981}
{Baldwin} J.~A.,  {Phillips} M.~M.,   {Terlevich} R.,  1981, \mn@doi [\pasp]
  {10.1086/130766}, \href {http://adsabs.harvard.edu/abs/1981PASP...93....5B}
  {93, 5}

\bibitem[\protect\citeauthoryear{{Begeman}}{{Begeman}}{1987}]{Begeman1987}
{Begeman} K.~G.,  1987, PhD thesis, , Kapteyn Institute, (1987)

\bibitem[\protect\citeauthoryear{{Bennert}, {Auger}, {Treu}, {Woo}  \&
  {Malkan}}{{Bennert} et~al.}{2011}]{Bennert2011}
{Bennert} V.~N.,  {Auger} M.~W.,  {Treu} T.,  {Woo} J.-H.,   {Malkan} M.~A.,
  2011, \mn@doi [\apj] {10.1088/0004-637X/742/2/107}, \href
  {http://adsabs.harvard.edu/abs/2011ApJ...742..107B} {742, 107}

\bibitem[\protect\citeauthoryear{{Blandford} \& {K{\"o}nigl}}{{Blandford} \&
  {K{\"o}nigl}}{1979}]{BK1979}
{Blandford} R.~D.,  {K{\"o}nigl} A.,  1979, \mn@doi [\apj] {10.1086/157262},
  \href {http://adsabs.harvard.edu/abs/1979ApJ...232...34B} {232, 34}

\bibitem[\protect\citeauthoryear{{Bosma}}{{Bosma}}{1978}]{Bosma1978}
{Bosma} A.,  1978, PhD thesis, PhD Thesis, Groningen Univ., (1978)

\bibitem[\protect\citeauthoryear{{Camenzind} \& {Courvoisier}}{{Camenzind} \&
  {Courvoisier}}{1983}]{CC1983}
{Camenzind} M.,  {Courvoisier} T.~J.-L.,  1983, \mn@doi [\apjl]
  {10.1086/183983}, \href {http://adsabs.harvard.edu/abs/1983ApJ...266L..83C}
  {266, L83}

\bibitem[\protect\citeauthoryear{{Cardelli}, {Clayton}  \& {Mathis}}{{Cardelli}
  et~al.}{1989}]{CCM1989}
{Cardelli} J.~A.,  {Clayton} G.~C.,   {Mathis} J.~S.,  1989, \mn@doi [\apj]
  {10.1086/167900}, \href {http://adsabs.harvard.edu/abs/1989ApJ...345..245C}
  {345, 245}

\bibitem[\protect\citeauthoryear{{Cecil}, {Dopita}, {Groves}, {Wilson},
  {Ferruit}, {P{\'e}contal}  \& {Binette}}{{Cecil} et~al.}{2002}]{Cecil2002}
{Cecil} G.,  {Dopita} M.~A.,  {Groves} B.,  {Wilson} A.~S.,  {Ferruit} P.,
  {P{\'e}contal} E.,   {Binette} L.,  2002, \mn@doi [\apj] {10.1086/338950},
  \href {http://adsabs.harvard.edu/abs/2002ApJ...568..627C} {568, 627}

\bibitem[\protect\citeauthoryear{{D'Agostino}, {Poetrodjojo}, {Ho}, {Groves},
  {Kewley}, {Madore}, {Rich}  \& {Seibert}}{{D'Agostino}
  et~al.}{2018}]{TYPHOONpaper}
{D'Agostino} J.~J.,  {Poetrodjojo} H.,  {Ho} I.-T.,  {Groves} B.,  {Kewley} L.,
   {Madore} B.~F.,  {Rich} J.,   {Seibert} M.,  2018, \mn@doi [\mnras]
  {10.1093/mnras/sty1676}, \href
  {http://adsabs.harvard.edu/abs/2018MNRAS.479.4907D} {479, 4907}

\bibitem[\protect\citeauthoryear{{D'Agostino}, {Kewley}, {Groves}, {Medling},
  {Dopita}  \& {Thomas}}{{D'Agostino} et~al.}{2019a}]{3dletter}
{D'Agostino} J.~J.,  {Kewley} L.~J.,  {Groves} B.~A.,  {Medling} A.,  {Dopita}
  M.~A.,   {Thomas} A.~D.,  2019a, \mn@doi [\mnras] {10.1093/mnrasl/slz028},
  \href {https://ui.adsabs.harvard.edu/abs/2019MNRAS.485L..38D} {485, L38}

\bibitem[\protect\citeauthoryear{{D'Agostino}, {Kewley}, {Groves}, {Byler},
  {Sutherland}, {Nicholls}, {Leitherer}  \& {Stanway}}{{D'Agostino}
  et~al.}{2019b}]{gridpaper}
{D'Agostino} J.~J.,  {Kewley} L.~J.,  {Groves} B.,  {Byler} N.,  {Sutherland}
  R.~S.,  {Nicholls} D.,  {Leitherer} C.,   {Stanway} E.~R.,  2019b, \mn@doi
  [\apj] {10.3847/1538-4357/ab1d5e}, \href
  {https://ui.adsabs.harvard.edu/abs/2019ApJ...878....2D} {878, 2}

\bibitem[\protect\citeauthoryear{{Davies}, {M{\"u}ller S{\'a}nchez}, {Genzel},
  {Tacconi}, {Hicks}, {Friedrich}  \& {Sternberg}}{{Davies}
  et~al.}{2007}]{Davies2007}
{Davies} R.~I.,  {M{\"u}ller S{\'a}nchez} F.,  {Genzel} R.,  {Tacconi} L.~J.,
  {Hicks} E.~K.~S.,  {Friedrich} S.,   {Sternberg} A.,  2007, \mn@doi [\apj]
  {10.1086/523032}, \href {http://adsabs.harvard.edu/abs/2007ApJ...671.1388D}
  {671, 1388}

\bibitem[\protect\citeauthoryear{{Davies}, {Rich}, {Kewley}  \&
  {Dopita}}{{Davies} et~al.}{2014a}]{Davies2014a}
{Davies} R.~L.,  {Rich} J.~A.,  {Kewley} L.~J.,   {Dopita} M.~A.,  2014a,
  \mn@doi [\mnras] {10.1093/mnras/stu234}, \href
  {http://adsabs.harvard.edu/abs/2014MNRAS.439.3835D} {439, 3835}

\bibitem[\protect\citeauthoryear{{Davies}, {Kewley}, {Ho}  \&
  {Dopita}}{{Davies} et~al.}{2014b}]{Davies2014b}
{Davies} R.~L.,  {Kewley} L.~J.,  {Ho} I.-T.,   {Dopita} M.~A.,  2014b, \mn@doi
  [\mnras] {10.1093/mnras/stu1740}, \href
  {http://adsabs.harvard.edu/abs/2014MNRAS.444.3961D} {444, 3961}

\bibitem[\protect\citeauthoryear{{Davies} et~al.,}{{Davies}
  et~al.}{2016}]{Davies2016}
{Davies} R.~L.,  et~al., 2016, \mn@doi [\mnras] {10.1093/mnras/stw1754}, \href
  {http://adsabs.harvard.edu/abs/2016MNRAS.462.1616D} {462, 1616}

\bibitem[\protect\citeauthoryear{{Davies} et~al.,}{{Davies}
  et~al.}{2017}]{Davies2017}
{Davies} R.~L.,  et~al., 2017, \mn@doi [\mnras] {10.1093/mnras/stx1559}, \href
  {http://adsabs.harvard.edu/abs/2017MNRAS.470.4974D} {470, 4974}

\bibitem[\protect\citeauthoryear{{Di Teodoro} \& {Fraternali}}{{Di Teodoro} \&
  {Fraternali}}{2015}]{Bbarolo}
{Di Teodoro} E.~M.,  {Fraternali} F.,  2015, \mn@doi [\mnras]
  {10.1093/mnras/stv1213}, \href
  {http://adsabs.harvard.edu/abs/2015MNRAS.451.3021D} {451, 3021}

\bibitem[\protect\citeauthoryear{{Dopita}}{{Dopita}}{1995}]{Dopita1995}
{Dopita} M.~A.,  1995, \mn@doi [\apss] {10.1007/BF00627353}, \href
  {https://ui.adsabs.harvard.edu/#abs/1995Ap&SS.233..215D} {233, 215}

\bibitem[\protect\citeauthoryear{{Dopita} \& {Sutherland}}{{Dopita} \&
  {Sutherland}}{2003}]{ADU}
{Dopita} M.~A.,  {Sutherland} R.~S.,  2003, {Astrophysics of the diffuse
  universe}

\bibitem[\protect\citeauthoryear{{Dopita}, {Groves}, {Sutherland}, {Binette}
  \& {Cecil}}{{Dopita} et~al.}{2002a}]{Dopita2002b}
{Dopita} M.~A.,  {Groves} B.~A.,  {Sutherland} R.~S.,  {Binette} L.,   {Cecil}
  G.,  2002a, \mn@doi [\apj] {10.1086/340429}, \href
  {http://adsabs.harvard.edu/abs/2002ApJ...572..753D} {572, 753}

\bibitem[\protect\citeauthoryear{{Dopita}, {Groves}, {Sutherland}, {Binette}
  \& {Cecil}}{{Dopita} et~al.}{2002b}]{Dopita2002}
{Dopita} M.~A.,  {Groves} B.~A.,  {Sutherland} R.~S.,  {Binette} L.,   {Cecil}
  G.,  2002b, \mn@doi [\apj] {10.1086/340429}, \href
  {http://adsabs.harvard.edu/abs/2002ApJ...572..753D} {572, 753}

\bibitem[\protect\citeauthoryear{{Dopita}, {Hart}, {McGregor}, {Oates},
  {Bloxham}  \& {Jones}}{{Dopita} et~al.}{2007}]{wifes1}
{Dopita} M.,  {Hart} J.,  {McGregor} P.,  {Oates} P.,  {Bloxham} G.,   {Jones}
  D.,  2007, \mn@doi [\apss] {10.1007/s10509-007-9510-z}, \href
  {http://adsabs.harvard.edu/abs/2007Ap%26SS.310..255D} {310, 255}

\bibitem[\protect\citeauthoryear{{Dopita} et~al.,}{{Dopita}
  et~al.}{2010}]{wifes2}
{Dopita} M.,  et~al., 2010, \mn@doi [\apss] {10.1007/s10509-010-0335-9}, \href
  {http://adsabs.harvard.edu/abs/2010Ap%26SS.327..245D} {327, 245}

\bibitem[\protect\citeauthoryear{{Dopita} et~al.,}{{Dopita}
  et~al.}{2015a}]{DopitaS7}
{Dopita} M.~A.,  et~al., 2015a, \mn@doi [\apjs] {10.1088/0067-0049/217/1/12},
  \href {http://adsabs.harvard.edu/abs/2015ApJS..217...12D} {217, 12}

\bibitem[\protect\citeauthoryear{{Dopita} et~al.,}{{Dopita}
  et~al.}{2015b}]{Dopita2015}
{Dopita} M.~A.,  et~al., 2015b, \mn@doi [\apj] {10.1088/0004-637X/801/1/42},
  \href {https://ui.adsabs.harvard.edu/#abs/2015ApJ...801...42D} {801, 42}

\bibitem[\protect\citeauthoryear{{Esquej} et~al.,}{{Esquej}
  et~al.}{2014}]{Esquej2014}
{Esquej} P.,  et~al., 2014, \mn@doi [\apj] {10.1088/0004-637X/780/1/86}, \href
  {http://adsabs.harvard.edu/abs/2014ApJ...780...86E} {780, 86}

\bibitem[\protect\citeauthoryear{{Ferland} \& {Mushotzky}}{{Ferland} \&
  {Mushotzky}}{1982}]{FM1982}
{Ferland} G.~J.,  {Mushotzky} R.~F.,  1982, \mn@doi [\apj] {10.1086/160448},
  \href {http://adsabs.harvard.edu/abs/1982ApJ...262..564F} {262, 564}

\bibitem[\protect\citeauthoryear{{Ferrarese} \& {Merritt}}{{Ferrarese} \&
  {Merritt}}{2000}]{FM2000}
{Ferrarese} L.,  {Merritt} D.,  2000, \mn@doi [\apjl] {10.1086/312838}, \href
  {http://adsabs.harvard.edu/abs/2000ApJ...539L...9F} {539, L9}

\bibitem[\protect\citeauthoryear{{Freitas} et~al.,}{{Freitas}
  et~al.}{2018}]{Freitas2018}
{Freitas} I.~C.,  et~al., 2018, \mn@doi [\mnras] {10.1093/mnras/sty303}, \href
  {https://ui.adsabs.harvard.edu/abs/2018MNRAS.476.2760F} {476, 2760}

\bibitem[\protect\citeauthoryear{{Gandhi}}{{Gandhi}}{2005}]{Gandhi2005}
{Gandhi} P.,  2005, Asian Journal of Physics, \href
  {http://adsabs.harvard.edu/abs/2005AsJPh..13...90G} {13, 90}

\bibitem[\protect\citeauthoryear{{Garc{\'{\i}}a-Burillo}
  et~al.,}{{Garc{\'{\i}}a-Burillo} et~al.}{2014}]{GB2014}
{Garc{\'{\i}}a-Burillo} S.,  et~al., 2014, \mn@doi [\aap]
  {10.1051/0004-6361/201423843}, \href
  {http://adsabs.harvard.edu/abs/2014A%26A...567A.125G} {567, A125}

\bibitem[\protect\citeauthoryear{{Garc{\'{\i}}a-Burillo}
  et~al.,}{{Garc{\'{\i}}a-Burillo} et~al.}{2016}]{GB2016}
{Garc{\'{\i}}a-Burillo} S.,  et~al., 2016, \mn@doi [\apjl]
  {10.3847/2041-8205/823/1/L12}, \href
  {http://adsabs.harvard.edu/abs/2016ApJ...823L..12G} {823, L12}

\bibitem[\protect\citeauthoryear{{Garc{\'{\i}}a-Burillo}
  et~al.,}{{Garc{\'{\i}}a-Burillo} et~al.}{2017}]{GB2017}
{Garc{\'{\i}}a-Burillo} S.,  et~al., 2017, \mn@doi [\aap]
  {10.1051/0004-6361/201731862}, \href
  {http://adsabs.harvard.edu/abs/2017A%26A...608A..56G} {608, A56}

\bibitem[\protect\citeauthoryear{{Gebhardt} et~al.,}{{Gebhardt}
  et~al.}{2000}]{Gebhardt2000}
{Gebhardt} K.,  et~al., 2000, \mn@doi [\apjl] {10.1086/312840}, \href
  {http://adsabs.harvard.edu/abs/2000ApJ...539L..13G} {539, L13}

\bibitem[\protect\citeauthoryear{{Gray} \& {Scannapieco}}{{Gray} \&
  {Scannapieco}}{2017}]{GS2017}
{Gray} W.~J.,  {Scannapieco} E.,  2017, \mn@doi [\apj]
  {10.3847/1538-4357/aa9121}, \href
  {http://adsabs.harvard.edu/abs/2017ApJ...849..132G} {849, 132}

\bibitem[\protect\citeauthoryear{{G{\"u}ltekin} et~al.,}{{G{\"u}ltekin}
  et~al.}{2009}]{Gultekin2009}
{G{\"u}ltekin} K.,  et~al., 2009, \mn@doi [\apj] {10.1088/0004-637X/698/1/198},
  \href {http://adsabs.harvard.edu/abs/2009ApJ...698..198G} {698, 198}

\bibitem[\protect\citeauthoryear{{Hampton} et~al.,}{{Hampton}
  et~al.}{2017}]{Hampton2017}
{Hampton} E.~J.,  et~al., 2017, \mn@doi [\mnras] {10.1093/mnras/stx1413}, \href
  {http://adsabs.harvard.edu/abs/2017MNRAS.470.3395H} {470, 3395}

\bibitem[\protect\citeauthoryear{{Heckman}, {Armus}  \& {Miley}}{{Heckman}
  et~al.}{1987}]{Heckman1987}
{Heckman} T.~M.,  {Armus} L.,   {Miley} G.~K.,  1987, \mn@doi [\aj]
  {10.1086/114310}, \href {http://adsabs.harvard.edu/abs/1987AJ.....93..276H}
  {93, 276}

\bibitem[\protect\citeauthoryear{{Ho} et~al.,}{{Ho} et~al.}{2014}]{Ho2014}
{Ho} I.-T.,  et~al., 2014, \mn@doi [\mnras] {10.1093/mnras/stu1653}, \href
  {http://adsabs.harvard.edu/abs/2014MNRAS.444.3894H} {444, 3894}

\bibitem[\protect\citeauthoryear{{Ho} et~al.,}{{Ho} et~al.}{2016}]{lzifu}
{Ho} I.-T.,  et~al., 2016, \mn@doi [\apss] {10.1007/s10509-016-2865-2}, \href
  {http://adsabs.harvard.edu/abs/2016Ap%26SS.361..280H} {361, 280}

\bibitem[\protect\citeauthoryear{{Kauffmann} et~al.,}{{Kauffmann}
  et~al.}{2003}]{Kauffmann2003}
{Kauffmann} G.,  et~al., 2003, \mn@doi [\mnras]
  {10.1111/j.1365-2966.2003.07154.x}, \href
  {http://adsabs.harvard.edu/abs/2003MNRAS.346.1055K} {346, 1055}

\bibitem[\protect\citeauthoryear{{Kennicutt}, {Tamblyn}  \&
  {Congdon}}{{Kennicutt} et~al.}{1994}]{Kennicutt1994}
{Kennicutt} Jr. R.~C.,  {Tamblyn} P.,   {Congdon} C.~E.,  1994, \mn@doi [\apj]
  {10.1086/174790}, \href {http://adsabs.harvard.edu/abs/1994ApJ...435...22K}
  {435, 22}

\bibitem[\protect\citeauthoryear{{Kewley} \& {Dopita}}{{Kewley} \&
  {Dopita}}{2002}]{KD2002}
{Kewley} L.~J.,  {Dopita} M.~A.,  2002, \mn@doi [\apjs] {10.1086/341326}, \href
  {http://adsabs.harvard.edu/abs/2002ApJS..142...35K} {142, 35}

\bibitem[\protect\citeauthoryear{{Kewley} \& {Ellison}}{{Kewley} \&
  {Ellison}}{2008}]{KE2008}
{Kewley} L.~J.,  {Ellison} S.~L.,  2008, \mn@doi [\apj] {10.1086/587500}, \href
  {http://adsabs.harvard.edu/abs/2008ApJ...681.1183K} {681, 1183}

\bibitem[\protect\citeauthoryear{{Kewley}, {Dopita}, {Sutherland}, {Heisler}
  \& {Trevena}}{{Kewley} et~al.}{2001}]{Kewley2001}
{Kewley} L.~J.,  {Dopita} M.~A.,  {Sutherland} R.~S.,  {Heisler} C.~A.,
  {Trevena} J.,  2001, \mn@doi [\apj] {10.1086/321545}, \href
  {http://adsabs.harvard.edu/abs/2001ApJ...556..121K} {556, 121}

\bibitem[\protect\citeauthoryear{{Kewley}, {Groves}, {Kauffmann}  \&
  {Heckman}}{{Kewley} et~al.}{2006}]{Kewley2006}
{Kewley} L.~J.,  {Groves} B.,  {Kauffmann} G.,   {Heckman} T.,  2006, \mn@doi
  [\mnras] {10.1111/j.1365-2966.2006.10859.x}, \href
  {http://adsabs.harvard.edu/abs/2006MNRAS.372..961K} {372, 961}

\bibitem[\protect\citeauthoryear{{Kewley}, {Dopita}, {Leitherer}, {Dav{\'e}},
  {Yuan}, {Allen}, {Groves}  \& {Sutherland}}{{Kewley}
  et~al.}{2013a}]{Kewley2013a}
{Kewley} L.~J.,  {Dopita} M.~A.,  {Leitherer} C.,  {Dav{\'e}} R.,  {Yuan} T.,
  {Allen} M.,  {Groves} B.,   {Sutherland} R.,  2013a, \mn@doi [\apj]
  {10.1088/0004-637X/774/2/100}, \href
  {http://adsabs.harvard.edu/abs/2013ApJ...774..100K} {774, 100}

\bibitem[\protect\citeauthoryear{{Kewley}, {Maier}, {Yabe}, {Ohta}, {Akiyama},
  {Dopita}  \& {Yuan}}{{Kewley} et~al.}{2013b}]{Kewley2013b}
{Kewley} L.~J.,  {Maier} C.,  {Yabe} K.,  {Ohta} K.,  {Akiyama} M.,  {Dopita}
  M.~A.,   {Yuan} T.,  2013b, \mn@doi [\apjl] {10.1088/2041-8205/774/1/L10},
  \href {http://adsabs.harvard.edu/abs/2013ApJ...774L..10K} {774, L10}

\bibitem[\protect\citeauthoryear{{Kraemer}, {Trippe}, {Crenshaw},
  {Mel{\'e}ndez}, {Schmitt}  \& {Fischer}}{{Kraemer}
  et~al.}{2009}]{Kraemer2009}
{Kraemer} S.~B.,  {Trippe} M.~L.,  {Crenshaw} D.~M.,  {Mel{\'e}ndez} M.,
  {Schmitt} H.~R.,   {Fischer} T.~C.,  2009, \mn@doi [\apj]
  {10.1088/0004-637X/698/1/106}, \href
  {http://adsabs.harvard.edu/abs/2009ApJ...698..106K} {698, 106}

\bibitem[\protect\citeauthoryear{{MacAlpine}}{{MacAlpine}}{1986}]{Macalpine1986}
{MacAlpine} G.~M.,  1986, \mn@doi [\pasp] {10.1086/131734}, \href
  {http://adsabs.harvard.edu/abs/1986PASP...98..134M} {98, 134}

\bibitem[\protect\citeauthoryear{{Magorrian} et~al.,}{{Magorrian}
  et~al.}{1998}]{Magorrian1998}
{Magorrian} J.,  et~al., 1998, \mn@doi [\aj] {10.1086/300353}, \href
  {http://adsabs.harvard.edu/abs/1998AJ....115.2285M} {115, 2285}

\bibitem[\protect\citeauthoryear{{Marconi} \& {Hunt}}{{Marconi} \&
  {Hunt}}{2003}]{MH2003}
{Marconi} A.,  {Hunt} L.~K.,  2003, \mn@doi [\apjl] {10.1086/375804}, \href
  {http://adsabs.harvard.edu/abs/2003ApJ...589L..21M} {589, L21}

\bibitem[\protect\citeauthoryear{{Marinucci} et~al.,}{{Marinucci}
  et~al.}{2016}]{Marinucci2016}
{Marinucci} A.,  et~al., 2016, \mn@doi [\mnras] {10.1093/mnrasl/slv178}, \href
  {http://adsabs.harvard.edu/abs/2016MNRAS.456L..94M} {456, L94}

\bibitem[\protect\citeauthoryear{{McConnell} \& {Ma}}{{McConnell} \&
  {Ma}}{2013}]{MM2013}
{McConnell} N.~J.,  {Ma} C.-P.,  2013, \mn@doi [\apj]
  {10.1088/0004-637X/764/2/184}, \href
  {http://adsabs.harvard.edu/abs/2013ApJ...764..184M} {764, 184}

\bibitem[\protect\citeauthoryear{{Miller} \& {Antonucci}}{{Miller} \&
  {Antonucci}}{1983}]{MA1983}
{Miller} J.~S.,  {Antonucci} R.~R.~J.,  1983, \mn@doi [\apjl] {10.1086/184082},
  \href {http://adsabs.harvard.edu/abs/1983ApJ...271L...7M} {271, L7}

\bibitem[\protect\citeauthoryear{{Miller}, {Goodrich}  \& {Mathews}}{{Miller}
  et~al.}{1991}]{Miller1991}
{Miller} J.~S.,  {Goodrich} R.~W.,   {Mathews} W.~G.,  1991, \mn@doi [\apj]
  {10.1086/170406}, \href {http://adsabs.harvard.edu/abs/1991ApJ...378...47M}
  {378, 47}

\bibitem[\protect\citeauthoryear{{Molteni}, {Lanzafame}  \&
  {Chakrabarti}}{{Molteni} et~al.}{1994}]{Molteni1994}
{Molteni} D.,  {Lanzafame} G.,   {Chakrabarti} S.~K.,  1994, \mn@doi [\apj]
  {10.1086/173972}, \href {http://adsabs.harvard.edu/abs/1994ApJ...425..161M}
  {425, 161}

\bibitem[\protect\citeauthoryear{{Neustroev} \& {Borisov}}{{Neustroev} \&
  {Borisov}}{1998}]{NB1998}
{Neustroev} V.~V.,  {Borisov} N.~V.,  1998, \aap, \href
  {http://adsabs.harvard.edu/abs/1998A%26A...336L..73N} {336, L73}

\bibitem[\protect\citeauthoryear{{Osterbrock} \& {Martel}}{{Osterbrock} \&
  {Martel}}{1993}]{OM1993}
{Osterbrock} D.~E.,  {Martel} A.,  1993, \mn@doi [\apj] {10.1086/173102}, \href
  {http://adsabs.harvard.edu/abs/1993ApJ...414..552O} {414, 552}

\bibitem[\protect\citeauthoryear{{Pogge}}{{Pogge}}{1988}]{Pogge1988}
{Pogge} R.~W.,  1988, \mn@doi [\apj] {10.1086/166309}, \href
  {http://adsabs.harvard.edu/abs/1988ApJ...328..519P} {328, 519}

\bibitem[\protect\citeauthoryear{{Rafferty}, {Brandt}, {Alexander}, {Xue},
  {Bauer}, {Lehmer}, {Luo}  \& {Papovich}}{{Rafferty}
  et~al.}{2011}]{Rafferty2011}
{Rafferty} D.~A.,  {Brandt} W.~N.,  {Alexander} D.~M.,  {Xue} Y.~Q.,  {Bauer}
  F.~E.,  {Lehmer} B.~D.,  {Luo} B.,   {Papovich} C.,  2011, \mn@doi [\apj]
  {10.1088/0004-637X/742/1/3}, \href
  {http://adsabs.harvard.edu/abs/2011ApJ...742....3R} {742, 3}

\bibitem[\protect\citeauthoryear{{Rich}, {Dopita}, {Kewley}  \& {Rupke}}{{Rich}
  et~al.}{2010}]{Rich2010}
{Rich} J.~A.,  {Dopita} M.~A.,  {Kewley} L.~J.,   {Rupke} D.~S.~N.,  2010,
  \mn@doi [\apj] {10.1088/0004-637X/721/1/505}, \href
  {http://adsabs.harvard.edu/abs/2010ApJ...721..505R} {721, 505}

\bibitem[\protect\citeauthoryear{{Rich}, {Kewley}  \& {Dopita}}{{Rich}
  et~al.}{2011}]{RKD2011}
{Rich} J.~A.,  {Kewley} L.~J.,   {Dopita} M.~A.,  2011, \mn@doi [\apj]
  {10.1088/0004-637X/734/2/87}, \href
  {http://adsabs.harvard.edu/abs/2011ApJ...734...87R} {734, 87}

\bibitem[\protect\citeauthoryear{{Rich}, {Kewley}  \& {Dopita}}{{Rich}
  et~al.}{2014}]{RKD2014}
{Rich} J.~A.,  {Kewley} L.~J.,   {Dopita} M.~A.,  2014, \mn@doi [\apjl]
  {10.1088/2041-8205/781/1/L12}, \href
  {http://adsabs.harvard.edu/abs/2014ApJ...781L..12R} {781, L12}

\bibitem[\protect\citeauthoryear{{Rupke} \& {Veilleux}}{{Rupke} \&
  {Veilleux}}{2011}]{Rupke2011}
{Rupke} D.~S.~N.,  {Veilleux} S.,  2011, \mn@doi [\apjl]
  {10.1088/2041-8205/729/2/L27}, \href
  {http://adsabs.harvard.edu/abs/2011ApJ...729L..27R} {729, L27}

\bibitem[\protect\citeauthoryear{{Rupke} \& {Veilleux}}{{Rupke} \&
  {Veilleux}}{2013}]{Rupke2013}
{Rupke} D.~S.~N.,  {Veilleux} S.,  2013, \mn@doi [\apj]
  {10.1088/0004-637X/768/1/75}, \href
  {http://adsabs.harvard.edu/abs/2013ApJ...768...75R} {768, 75}

\bibitem[\protect\citeauthoryear{{Rupke}, {Veilleux}  \& {Sanders}}{{Rupke}
  et~al.}{2005}]{Rupke2005}
{Rupke} D.~S.,  {Veilleux} S.,   {Sanders} D.~B.,  2005, \mn@doi [\apjs]
  {10.1086/432889}, \href {http://adsabs.harvard.edu/abs/2005ApJS..160..115R}
  {160, 115}

\bibitem[\protect\citeauthoryear{{Schoniger} \& {Sofue}}{{Schoniger} \&
  {Sofue}}{1994}]{1068dist}
{Schoniger} F.,  {Sofue} Y.,  1994, \aap, \href
  {https://ui.adsabs.harvard.edu/abs/1994A&A...283...21S} {283, 21}

\bibitem[\protect\citeauthoryear{{Sponholz} \& {Molteni}}{{Sponholz} \&
  {Molteni}}{1994}]{SM1994}
{Sponholz} H.,  {Molteni} D.,  1994, \mn@doi [\mnras]
  {10.1093/mnras/271.1.233}, \href
  {http://adsabs.harvard.edu/abs/1994MNRAS.271..233S} {271}

\bibitem[\protect\citeauthoryear{{Spruit}}{{Spruit}}{1987}]{Spruit1987}
{Spruit} H.~C.,  1987, \aap, \href
  {http://adsabs.harvard.edu/abs/1987A%26A...184..173S} {184, 173}

\bibitem[\protect\citeauthoryear{{Storchi-Bergmann}, {Riffel}, {Riffel},
  {Diniz}, {Borges Vale}  \& {McGregor}}{{Storchi-Bergmann}
  et~al.}{2012}]{SB2012}
{Storchi-Bergmann} T.,  {Riffel} R.~A.,  {Riffel} R.,  {Diniz} M.~R.,  {Borges
  Vale} T.,   {McGregor} P.~J.,  2012, \mn@doi [\apj]
  {10.1088/0004-637X/755/2/87}, \href
  {http://adsabs.harvard.edu/abs/2012ApJ...755...87S} {755, 87}

\bibitem[\protect\citeauthoryear{{Sutherland}, {Dopita}, {Binette}  \&
  {Groves}}{{Sutherland} et~al.}{2018}]{mappingsv}
{Sutherland} R.,  {Dopita} M.,  {Binette} L.,   {Groves} B.,  2018, {MAPPINGS
  V: Astrophysical plasma modeling code}, Astrophysics Source Code Library
  (\mn@eprint {ascl} {1807.005})

\bibitem[\protect\citeauthoryear{{Tacconi}, {Genzel}, {Blietz}, {Cameron},
  {Harris}  \& {Madden}}{{Tacconi} et~al.}{1994}]{Tacconi1994}
{Tacconi} L.~J.,  {Genzel} R.,  {Blietz} M.,  {Cameron} M.,  {Harris} A.~I.,
  {Madden} S.,  1994, \mn@doi [\apjl] {10.1086/187344}, \href
  {http://adsabs.harvard.edu/abs/1994ApJ...426L..77T} {426, 77}

\bibitem[\protect\citeauthoryear{{Taylor}, {Tadhunter}  \& {Robinson}}{{Taylor}
  et~al.}{2003}]{Taylor2003}
{Taylor} M.~D.,  {Tadhunter} C.~N.,   {Robinson} T.~G.,  2003, \mn@doi [\mnras]
  {10.1046/j.1365-8711.2003.06615.x}, \href
  {http://adsabs.harvard.edu/abs/2003MNRAS.342..995T} {342, 995}

\bibitem[\protect\citeauthoryear{{Thomas} et~al.,}{{Thomas} et~al.}{2017}]{S7}
{Thomas} A.~D.,  et~al., 2017, \mn@doi [\apjs] {10.3847/1538-4365/aa855a},
  \href {http://adsabs.harvard.edu/abs/2017ApJS..232...11T} {232, 11}

\bibitem[\protect\citeauthoryear{{Thronson} Jr. et~al.,}{{Thronson}
  et~al.}{1989}]{Thronson1989}
{Thronson} Jr. H.~A.,  et~al., 1989, \mn@doi [\apj] {10.1086/167693}, \href
  {http://adsabs.harvard.edu/abs/1989ApJ...343..158T} {343, 158}

\bibitem[\protect\citeauthoryear{{Tremaine} et~al.,}{{Tremaine}
  et~al.}{2002}]{Tremaine2002}
{Tremaine} S.,  et~al., 2002, \mn@doi [\apj] {10.1086/341002}, \href
  {http://adsabs.harvard.edu/abs/2002ApJ...574..740T} {574, 740}

\bibitem[\protect\citeauthoryear{{Vaona}, {Ciroi}, {Di Mille}, {Cracco}, {La
  Mura}  \& {Rafanelli}}{{Vaona} et~al.}{2012}]{Vaona2012}
{Vaona} L.,  {Ciroi} S.,  {Di Mille} F.,  {Cracco} V.,  {La Mura} G.,
  {Rafanelli} P.,  2012, \mn@doi [\mnras] {10.1111/j.1365-2966.2012.22060.x},
  \href {http://adsabs.harvard.edu/abs/2012MNRAS.427.1266V} {427, 1266}

\bibitem[\protect\citeauthoryear{{Veilleux} \& {Osterbrock}}{{Veilleux} \&
  {Osterbrock}}{1987}]{VO1987}
{Veilleux} S.,  {Osterbrock} D.~E.,  1987, \mn@doi [\apjs] {10.1086/191166},
  \href {http://adsabs.harvard.edu/abs/1987ApJS...63..295V} {63, 295}

\bibitem[\protect\citeauthoryear{{Wilson} \& {Raymond}}{{Wilson} \&
  {Raymond}}{1999}]{WR1999}
{Wilson} A.~S.,  {Raymond} J.~C.,  1999, \mn@doi [\apj] {10.1086/311923}, \href
  {https://ui.adsabs.harvard.edu/#abs/1999ApJ...513L.115W} {513, L115}

\bibitem[\protect\citeauthoryear{{Woltjer}}{{Woltjer}}{1959}]{Woltjer1959}
{Woltjer} L.,  1959, \mn@doi [\apj] {10.1086/146694}, \href
  {http://adsabs.harvard.edu/abs/1959ApJ...130...38W} {130, 38}

\bibitem[\protect\citeauthoryear{{Young}, {Wilson}  \& {Shopbell}}{{Young}
  et~al.}{2001}]{Young2001}
{Young} A.~J.,  {Wilson} A.~S.,   {Shopbell} P.~L.,  2001, \mn@doi [\apj]
  {10.1086/321561}, \href {http://adsabs.harvard.edu/abs/2001ApJ...556....6Y}
  {556, 6}

\bibitem[\protect\citeauthoryear{{Yuan}, {Kewley}  \& {Sanders}}{{Yuan}
  et~al.}{2010}]{YKS2010}
{Yuan} T.-T.,  {Kewley} L.~J.,   {Sanders} D.~B.,  2010, \mn@doi [\apj]
  {10.1088/0004-637X/709/2/884}, \href
  {http://adsabs.harvard.edu/abs/2010ApJ...709..884Y} {709, 884}

\bibitem[\protect\citeauthoryear{{Zubovas} \& {King}}{{Zubovas} \&
  {King}}{2012}]{ZK2012}
{Zubovas} K.,  {King} A.~R.,  2012, in {Chartas} G.,  {Hamann} F.,   {Leighly}
  K.~M.,  eds,  Astronomical Society of the Pacific Conference Series Vol. 460,
  AGN Winds in Charleston. p.~235 (\mn@eprint {arXiv} {1201.3540})

\bibitem[\protect\citeauthoryear{{de Vaucouleurs}, {de Vaucouleurs}, {Corwin},
  {Buta}, {Paturel}  \& {Fouqu{\'e}}}{{de Vaucouleurs} et~al.}{1991}]{dV1991}
{de Vaucouleurs} G.,  {de Vaucouleurs} A.,  {Corwin} Jr. H.~G.,  {Buta} R.~J.,
  {Paturel} G.,   {Fouqu{\'e}} P.,  1991, {Third Reference Catalogue of Bright
  Galaxies. Volume I: Explanations and references. Volume II: Data for galaxies
  between 0$^{h}$ and 12$^{h}$. Volume III: Data for galaxies between 12$^{h}$
  and 24$^{h}$.}

\makeatother
\end{thebibliography}







\bsp	
\label{lastpage}
\end{document}